\documentclass[draftcls, one column]{IEEEtran}
\usepackage{amsmath,amssymb}
\usepackage{graphicx}
\usepackage{mathrsfs}
\usepackage{tikz}
\usepackage{booktabs}
\usepackage{enumerate}
\usepackage{psfrag}	
\usepackage{array}
\usepackage{multirow,hhline}
\usepackage{exscale}
\usepackage{color}
\usepackage{colortbl}
\usepackage{epsfig,subfigure}
\usepackage{cite}
\usepackage{amsthm}
\usepackage{hyperref}
\usepackage{dsfont}
\usepackage{soul}	

\newtheorem{theorem}{Theorem}
\newtheorem{lemma}{Lemma}
\newtheorem{remark}{Remark}

\newtheorem{proposition}{Proposition}

\renewcommand{\sc}[0]{\ensuremath{\mathsf{c}}}

\newcommand{\snr}[0]{\ensuremath{\mathsf{SNR}}}

\graphicspath{{./fig/}}

\usepackage{tikz}
\usepackage{pgfplots}
\usepackage{pgfplotstable}
\usepackage{rotate}

\definecolor{rubblue}{cmyk}{1,0.5,0,0.6}
\definecolor{rubgreen}{cmyk}{0.5,0,1,0}
\definecolor{rubgray}{cmyk}{0.03,0.03,0.03,0.1}

\usepgfplotslibrary{units}

\usetikzlibrary{%
patterns,%
calc,%
fit,%
arrows,%
plotmarks,%
shadows,%
chains,%
shapes%
}

\tikzset{>=latex'} 
\tikzstyle{every picture}+=[remember picture] 
\pgfdeclarelayer{background}
\pgfdeclarelayer{foreground}
\pgfsetlayers{background,main,foreground}

\tikzstyle{blueblock}=[draw=rubblue, rectangle, thick, drop shadow, minimum width=20mm, minimum height=8mm,fill=rubblue!20, text width=20mm, text centered]
\tikzstyle{bluebox}=[draw=rubblue, rectangle, thick, drop shadow, minimum width=8mm, minimum height=8mm,fill=rubblue!20, text width=8mm, text centered]
\tikzstyle{greenblock}=[draw=rubgreen, rectangle, thick, drop shadow, minimum width=20mm, minimum height=8mm,fill=rubgreen!20, text width=20mm, text centered]
\tikzstyle{dot} = [draw, circle, minimum size=0.1pt,scale=0.1,fill=black,black]
\tikzstyle{reddot} = [draw, circle, minimum size=0.2pt,scale=0.3,fill=Red,Red]
\tikzstyle{blackdot} = [draw, circle, minimum size=0.2pt,scale=0.7,fill=black,black]
\tikzstyle{sum} = [drop shadow, draw=rubblue, thick, fill=rubblue!20, circle]
\tikzstyle{whitedot} = [draw, circle, minimum size=0.2pt, scale=0.8, fill=white, thin]
\tikzstyle{relay} = [blueblock, minimum width=5mm, minimum height=20mm, text width=5mm, rounded corners=2pt]
\tikzstyle{relay2} = [blueblock, minimum width=5mm, minimum height=15mm, text width=5mm, rounded corners=2pt]
\tikzstyle{relay3} = [blueblock, minimum width=5mm, minimum height=25mm, text width=5mm, rounded corners=2pt]
\tikzstyle{relay4} = [blueblock, minimum width=5mm, minimum height=10mm, text width=5mm, rounded corners=2pt]
\tikzstyle{relay5} = [blueblock, minimum width=5mm, minimum height=50mm, text width=5mm, rounded corners=2pt]
\tikzstyle{relay6} = [blueblock, minimum width=5mm, minimum height=5mm, text width=5mm, rounded corners=2pt]
\tikzstyle{circgreen} = [draw, circle, inner sep=2pt, fill=rubgreen, drop shadow, thick]
\tikzstyle{circwhite} = [draw, circle, inner sep=2pt, fill=white, drop shadow, thick]
\tikzstyle{circdashed} = [draw, dashed, circle, inner sep=2pt, fill=rubgray, drop shadow, thick]
\tikzstyle{vertbox} = [rectangle, draw=rubblue, thick, rotate=90, text centered, minimum width=16.5mm, minimum height=8mm, text width=16.5mm, inner sep=0pt, fill=rubblue!20, drop shadow]
\tikzstyle{vertboxb} = [rectangle, draw=rubblue, thick, rotate=90, text centered, minimum width=16.5mm, minimum height=8mm, text width=16.5mm, fill=rubblue!20, drop shadow]
\tikzstyle{vertboxshort} = [rectangle, draw=rubblue, thick, rotate=90, text centered, minimum width=10mm, minimum height=8mm, text width=10mm, inner sep=0pt, fill=rubblue!20, drop shadow]
\tikzstyle{smalldotgreen} = [draw=rubgreen, circle, minimum size=0.2pt,scale=0.8,fill=rubgreen!20]
\tikzstyle{antenna} = [regular polygon, regular polygon sides=3, draw, shape border rotate=180, minimum size=0.2pt, scale=0.3]

\tikzstyle{poly} = [regular polygon, regular polygon sides=6, shape aspect=0.5, minimum width=1.5cm, minimum height=0.35cm, draw, dashed]

\tikzset{axis/.style={ultra thick, red, -latex, shorten <=-.5 cm, shorten >=-1 cm}}

\definecolor{cff9e00}{RGB}{255,158,0}
\definecolor{c4fff00}{RGB}{79,255,0}
\definecolor{cff0012}{RGB}{255,0,18}
\definecolor{c00c5ff}{RGB}{0,197,255}
\definecolor{c046f00}{RGB}{4,111,0}
\definecolor{c004b9d}{RGB}{0,75,157}

\newlength{\mylen}
\usepackage[outline]{contour}
\contourlength{1.5pt}

\begin{document}

\title{Three-Way Channels with Multiple Unicast Sessions: Capacity Approximation via Network Transformation}
\author{Anas~Chaaban, 
Henning~Maier, 
Aydin~Sezgin, 
and Rudolf Mathar
\thanks{A. Chaaban is with the Division of Computer, Electrical, and Mathematical Sciences and Engineering (CEMSE) at King Abdullah University of Science and Technology (KAUST), Thuwal, Saudi Arabia. Email: anas.chaaban@kaust.edu.sa.

H. Maier and R. Mathar are with the Institute for Theoretical Information Technology, RWTH Aachen University, 52056 Aachen, Germany. Email: \{maier, mathar\}@ti.rwth-aachen.de.

A. Sezgin is with the Institute of Digital Communication Systems, Ruhr-Universit\"at Bochum (RUB), Universit\"atsstrasse 150, 44780 Bochum, Germany. Email: aydin.sezgin@rub.de. The work of A. Sezgin was supported in part by the DFG, under grant SE 1697/5.}


\thanks{Parts of the paper have been presented in Allerton 2014 \cite{MaierChaabanMatharSezgin} and at the ISIT 2014 \cite{ChaabanMaierSezgin}.}

}



\maketitle

\begin{abstract}
A network of 3 nodes mutually communicating with each other is studied. This multi-way network is a suitable model for 3-user device-to-device communications. The main goal of this paper is to characterize the capacity region of the underlying Gaussian 3-way channel (3WC) within a constant gap. To this end, a capacity outer bound is derived using cut-set bounds and genie-aided bounds. For achievability, the 3WC is first transformed into an equivalent star-channel. This latter is then decomposed into a set of `successive' sub-channels, leading to a sub-channel allocation problem. Using backward decoding, interference neutralization, and known results on the capacity of the star-channel relying of physical-layer network coding, an achievable rate region for the 3WC is obtained. It is then shown that the achievable rate region is within a constant gap of the developed outer bound, leading to the desired capacity approximation. Interestingly, in contrast to the Gaussian two-way channel (TWC), adaptation is necessary in the 3WC. Furthermore, message splitting is another ingredient of the developed scheme for the 3WC which is not required in the TWC. The two setups are, however, similar in terms of their sum-capacity pre-log which is equal to 2. Finally, some interesting networks and their approximate capacities are recovered as special cases of the 3WC, such as the cooperative BC and MAC.
\end{abstract}


\section{Introduction}
\IEEEpubidadjcol
\IEEEPARstart{F}{uture} communication networks have to support much higher data-rates than today's networks~\cite{Evans_IoT}. Towards this goal, several new ideas are presently being investigated, such as full-duplex and multi-way communication.\cite{ChaabanSezginFnT}.

This simplest form of multi-way communication is the two-way channel (TWC) studied in \cite{Shannon_TWC,Han}, where two nodes communicate with each other in both directions. This can be established in a half-duplex (HD) or a full-duplex (FD) mode. HD is the common trend in today's communication systems due to its practical simplicity. An HD TWC can be modeled as two non-interacting point-to-point (P2P) channels, for which both theory and practice have matured since Shannon's work in~\cite{Shannon}. The draw-back of HD is its halved spectral efficiency; a communication bandwidth of $B$~Hz has to be segregated into two bands $B/2$~Hz each, one for each communication direction. 
FD operation can avoid such data-rate reduction.

In an FD TWC, communication in both directions takes place over the same bandwidth $B$. FD has witnessed important developments recently and reached practical maturity \cite{ChoiJain,HongBrand,SabharwalSchniterGuo}. The advantage of FD is that it doubles the data rate~\cite{Bharadia} in comparison to HD. For instance, the capacity of the FD Gaussian TWC is equal to that of two parallel P2P channels \cite{Han} operating over the whole resources available (time/frequency). From this point-of-view, FD multi-way communication is regarded as a promising candidate for boosting the performance of future networks.

An example of FD two-way operation is a device-to-device communications (D2D) scenario \cite{NaderializadehAvestimehr}. D2D has been proposed by several researchers as a potential component of future networks \cite{LayaWangWidaaZarate,AsadiWangMancuso}. It allows nearby users to communicate directly with limited or no involvement of the base-station (BS), in what is known as D2D communication with operator control (DC-OC) and D2D communication with device control (DC-DC), respectively~\cite{TehraniUysalYanikomeroglu}. D2D can offload some traffic from the BS in dense areas, especially if the D2D pair and the BS operate in-band. Consider a general scenario where a D2D pair and a BS share the same resources, and want to establish full message exchange.\footnote{Information exchange between the BS and the D2D nodes might encompass control signaling, or communication with a third party in another cell for example.} In its full generality, such a scenario can be modeled as a fully-connected 3-way network. This also models 3-user D2D communication operating in device relaying with device control (DR-DC) or operator control (DR-OC) mode \cite{TehraniUysalYanikomeroglu,ChaababSezginCrownCom}. 

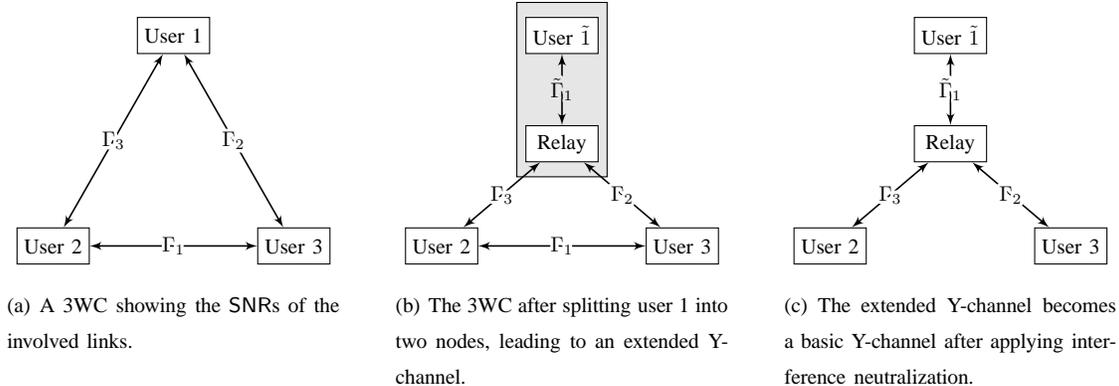
\begin{figure}[t]
\centering
\subfigure[A 3WC showing the $\snr$s of the involved links.]{
\begin{tikzpicture}[semithick,scale=0.8, every node/.style={scale=0.8}]
\node (u1) at (2,3.5) [rectangle, draw, thin, fill=white, minimum width=1.2cm, minimum height=.6cm, rotate=0] {User 1};
\node (u2) at (0,0) [rectangle, draw, thin, fill=white, minimum width=1.2cm, minimum height=.6cm, rotate=0] {User 2};
\node (u3) at (4,0) [rectangle, draw, thin, fill=white, minimum width=1.2cm, minimum height=.6cm, rotate=0] {User 3};

\draw[<->] (u1) to node[] {\contour{white}{$\Gamma_3$}} (u2);
\draw[<->] (u2) to node[] {\contour{white}{$\Gamma_1$}}  (u3);
\draw[<->] (u3) to node[] {\contour{white}{$\Gamma_2$}}  (u1);
\end{tikzpicture}
\label{Fig:DYTrans1}
}
\hspace{.5cm}
\subfigure[The 3WC after splitting user~1 into two nodes, leading to an extended Y-channel.]{
\begin{tikzpicture}[semithick,scale=0.8, every node/.style={scale=0.8}]
\node at (2,2.6) [rectangle, draw, thin, fill=gray!20, minimum width=1.5cm, minimum height=2.9cm, rotate=0] {};
\node (u1) at (2,3.5) [rectangle, draw, thin, fill=white, minimum width=1.2cm, minimum height=.6cm, rotate=0] {User $\tilde{1}$};
\node (r) at (2,1.7) [rectangle, draw, thin, fill=white, minimum width=1.2cm, minimum height=.6cm, rotate=0] {Relay};
\node (u2) at (0,0) [rectangle, draw, thin, fill=white, minimum width=1.2cm, minimum height=.6cm, rotate=0] {User 2};
\node (u3) at (4,0) [rectangle, draw, thin, fill=white, minimum width=1.2cm, minimum height=.6cm, rotate=0] {User 3};

\draw[<->] (r) to node[] {\contour{white}{$\Gamma_3$}} (u2);
\draw[<->] (u2) to node[] {\contour{white}{$\Gamma_1$}}  (u3);
\draw[<->] (u3) to node[] {\contour{white}{$\Gamma_2$}}  (r);
\draw[<->] (r) to node[] {\contour{gray!20}{$\tilde{\Gamma}_1$}}  (u1);
\end{tikzpicture}
\label{Fig:DYTrans2}
}
\hspace{.5cm}
\subfigure[The extended Y-channel becomes a basic Y-channel after applying interference neutralization.]{
\begin{tikzpicture}[semithick,scale=0.8, every node/.style={scale=0.8}]
\node (u1) at (2,3.5) [rectangle, draw, thin, fill=white, minimum width=1.2cm, minimum height=.6cm, rotate=0] {User $\tilde{1}$};
\node (r) at (2,1.7) [rectangle, draw, thin, fill=white, minimum width=1.2cm, minimum height=.6cm, rotate=0] {Relay};
\node (u2) at (0,0) [rectangle, draw, thin, fill=white, minimum width=1.2cm, minimum height=.6cm, rotate=0] {User 2};
\node (u3) at (4,0) [rectangle, draw, thin, fill=white, minimum width=1.2cm, minimum height=.6cm, rotate=0] {User 3};

\draw[<->] (r) to node[] {\contour{white}{$\Gamma_3$}} (u2);
\draw[<->] (u3) to node[] {\contour{white}{$\Gamma_2$}}  (r);
\draw[<->] (r) to node[] {\contour{white}{$\tilde{\Gamma}_1$}}  (u1);
\end{tikzpicture}
\label{Fig:DYTrans3}
}
\caption{Transforming the 3-way channel (3WC) into a $Y$-channel. The value of $\tilde{\Gamma}_1$ will be specified later.}
\label{Fig:DYTrans}
\end{figure}

The resulting 3-way channel (3WC) shown in Fig. \ref{Fig:DYTrans1} ($\Delta$-channel in \cite{ChaabanMaierSezgin}) is the main focus of this work. We consider a multiple uni-cast setting where each node sends a private message to each other node. The multi cast setting where each node sends a common message to the other nodes was considered in \cite{Ong}. The 3WC can be thought of as an extension of the TWC to 3 nodes. Other extensions that have been considered in the literature include two-way multiple access, broadcast, and interference channels \cite{SuhWangTse,VahidSuhAvestimehr,ChengDevroye,
ChengDevroye_Kpair}, and two-way relay networks \cite{ChengDevroyeLiu,NamChungLee_IT,NgoLarsson,
SezginAvestimehrKhajehnejadHassibi,GunduzYenerGoldsmithPoor_IT,
TianYenerMIMOMW}.

The 3WC combines several aspects of wireless communication like multiple-access, broadcasting, and most importantly relaying. Each node in a 3WC is simultaneously a source, a destination, and a relay that can be uni-directional \cite{CoverElgamal} and/or bi-directional \cite{OechteringBjelakovicSchnurrBoche, GunduzTuncelNayak, AvestimehrSezginTse}. Finding the best transmission scheme over this channel is interesting from both theoretical and practical perspectives. The main goal of this paper is thus \emph{characterizing the capacity region of this 3WC within a constant gap}. Instead of studying the 3WC directly, we resort to a detour by transforming the 3WC to a star-channel (Y-channel) \cite{LeeLimChun, ChaabanSezginAvestimehr_YC_SC} first (Fig. \ref{Fig:DYTrans3}). The Y-channel is a multi-way relay channel \cite{GunduzYenerGoldsmithPoor,
SezginAvestimehrKhajehnejadHassibi,
OngKellettJohnson,ZewailMohassebNafieElGamal} which differs from the 3WC in that information is exchanged indirectly via a relay node. Then, we derive results on the capacity of the 3WC from results on the capacity of the transformed Y-channel. This 3WC--Y-channel transformation (analogous with the $\Delta$--Y transformation) serves as a useful tool for studying the 3WC. It might also be beneficial for studying larger multi-way channels as well. 

Our approach can be summarized as follows:
\begin{itemize}
\item Deriving a capacity outer bound for the 3WC,
\item transforming the 3WC to a Y-channel, and
\item deriving an achievable rate region using a successive channel decomposition (SCD) similar to \cite{BreslerParekhTse_IT}.
\end{itemize}
The outer bound is derived using the cut-set bound \cite{CoverThomas} and novel genie-aided bounds. This outer bound is identical (within a constant gap) to the approximate capacity region of a Y-channel \cite{ChaabanSezgin_YC_Reg}. This motivates extending the transmission scheme of the Y-channel in \cite{ChaabanSezgin_YC_Reg} to the 3WC. Motivated by this, we transform the 3WC to a Y-channel (Fig. \ref{Fig:DYTrans2}). We then decompose this Y-channel into a set of sub-channels using the SCD, and apply a scheme for the basic Y-channel (Fig. \ref{Fig:DYTrans3}) over each sub-channel. This scheme involves physical-layer network-coding using nested-lattice codes \cite{NazerGastpar}.\footnote{The application of physical-layer network-coding using nested-lattice codes is common in this context, see \cite{VahidSuhAvestimehr,SezginAvestimehrKhajehnejadHassibi,
GunduzYenerGoldsmithPoor_IT} e.g.} The interference (see Fig. \ref{Fig:DYTrans2}) is resolved using backward decoding and interference neutralization. This scheme achieves the outer bound within a constant gap. The resulting transmission scheme and capacity region approximation can be readily applied to the aforementioned D2D scenarios for any mode of information exchange between the 3 nodes. 

It is worth to note the contrast between the TWC and the 3WC. In the Gaussian TWC, adaptation and message splitting are not necessary \cite{Han}. However, those are main ingredients in the developed transmission scheme for the 3WC. In fact, adaptation is necessary in the 3WC as shown in this paper. The TWC and the 3WC are, however, similar in terms of their sum-capacity pre-log which is equal to 2 as shown in \cite{ChaabanMaierSezgin}.

As a side contribution, we provide a general framework for applying the SCD to a Gaussian network with arbitrary topology. We note that the SCD appeared first, to the best of our knowledge, in \cite{BreslerParekhTse_IT,BreslerParekhTse} where it was described for an interference channel. It was used later on for other networks \cite{SridharanJafarianVishwanathJafarShamai,
ChaabanSezgin_IT_IRC,SahaBerry,
SezginAvestimehrKhajehnejadHassibi,VahidSuhAvestimehr}. Similar to the deterministic channel model \cite{AvestimehrDiggaviTse_IT}, the Q-ary channel decomposition \cite{JafarVishwanath}, and the lower-triangular decomposition \cite{NiesenMaddah}, the SCD decomposes a channel into a set of sub-channels, effectively reducing the rate achievability problem to a sub-channel allocation problem. This allocation can be tackled very efficiently by using linear programming methods. The main advantage is that the SCD is applicable directly for the Gaussian channel without any further assumptions, contrary to the other decompositions.


The rest of the paper is organized as follows. Section \ref{Sec:NotationAndSystemModel} introduces the notation of the paper and the system model of the 3WC, in addition to the Y-channel. Section \ref{Sec:Summary} presents the main result. Section \ref{Sec:DCUpperBounds} provides an outer bound for the 3WC. Section \ref{Sec:DCAchievability} present an achievable  rate region for the 3WC using the aforementioned transformation to a Y-channel, and proves the achievability of the outer bound within a constant gap. Section \ref{Sec:SpecialCases} discusses special cases of the 3WC that have been studied earlier in the literature, namely cooperative multiple-access and broadcast channels. We conclude in Section \ref{Sec:Conc}.

\section{Notation and System Model}
\label{Sec:NotationAndSystemModel}

\subsection{Notation}
Throughout the paper, we denote a sequence of symbols $(x(1),\cdots,x(n))$ by $x^{(n)}$, we denote $\frac{1}{2}\log(1+x)$ by  $C(x)$, and we use $\hat{C}(x)$ to denote $\max\{0,C(x-1)\}$ which is an approximation of $C(x)$ for large $x$. We write $X\sim\mathcal{N}(\mu,\sigma^2)$ to indicate that $X$ is Gaussian distributed with mean $\mu$ and variance $\sigma^2$, and we use i.i.d. to indicate that a random sequence has independent and identically distributed instances.

\subsection{System Model: 3-Way Channel}
\label{sec:sysmod-3way}
\begin{figure*}[t]
\centering
\subfigure[A $3$-way channel.]{
\begin{tikzpicture}[semithick,scale=.9]
\node (u1) at (2,3.5) [rectangle, draw, thin, fill=white, minimum width=1.2cm, minimum height=.6cm, rotate=0] {User 1};
\node (u2) at (0,0) [rectangle, draw, thin, fill=white, minimum width=1.2cm, minimum height=.6cm, rotate=0] {User 2};
\node (u3) at (4,0) [rectangle, draw, thin, fill=white, minimum width=1.2cm, minimum height=.6cm, rotate=0] {User 3};

\draw[<->] (u1) to node[] {\contour{white}{$h_3$}} (u2);
\draw[<->] (u2) to node[] {\contour{white}{$h_1$}}  (u3);
\draw[<->] (u3) to node[] {\contour{white}{$h_2$}}  (u1);

\draw[->] (u1.east) to ($(u1.east)+(.5,0)$);
\node at ($(u1.east)+(.9,.25)$) {$\hat{w}_{12}$};
\node at ($(u1.east)+(.9,-.25)$) {$\hat{w}_{13}$};
\draw[->] ($(u1.west)-(0.5,0)$) to (u1.west);
\node at ($(u1.west)-(.9,.25)$) {$w_{21}$};
\node at ($(u1.west)-(.9,-.25)$) {$w_{31}$};

\draw[->] ($(u2.south)$) to ($(u2.south)+(0,-.5)$);
\node at ($(u2.south)+(0,-.75)$) {$\hat{w}_{21},\hat{w}_{23}$};
\draw[->] ($(u2.west)-(0.5,0)$) to ($(u2.west)$);
\node at ($(u2.west)-(.9,.25)$) {$w_{12}$};
\node at ($(u2.west)-(.9,-.25)$) {$w_{32}$};

\draw[->] (u3.south) to ($(u3.south)+(0,-.5)$);
\node at ($(u3.south)+(0,-.7)$) {$\hat{w}_{31},\hat{w}_{32}$};
\draw[->] ($(u3.east)+(.5,0)$) to (u3.east);
\node at ($(u3.east)+(.9,.25)$) {$w_{13}$};
\node at ($(u3.east)+(.9,-.25)$) {$w_{23}$};
\end{tikzpicture}
\label{Fig:3WCModel}
}
\hspace{.3cm}
\subfigure[A Y-channel.]{
\begin{tikzpicture}[semithick,scale=.9]
\node (u1) at (0,3.5) [rectangle, draw, thin, fill=white, minimum width=1.2cm, minimum height=.6cm, rotate=0] {User 1};
\node (u2) at (2,0) [rectangle, draw, thin, fill=white, minimum width=1.2cm, minimum height=.6cm, rotate=0] {User 2};
\node (r) at (2,2) [rectangle, draw, thin, fill=white, minimum width=1.2cm, minimum height=.6cm, rotate=0] {Relay};
\node (u3) at (4,3.5) [rectangle, draw, thin, fill=white, minimum width=1.2cm, minimum height=.6cm, rotate=0] {User 3};

\draw[<->] (u1) to node[] {\contour{white}{$\tilde{h}_1$}}  (r);
\draw[<->] (u2) to node[] {\contour{white}{$\tilde{h}_2$}}  (r);
\draw[<->] (u3) to node[] {\contour{white}{$\tilde{h}_3$}}  (r);

\draw[->] (u1.north) to ($(u1.north)+(0,.5)$);
\node at ($(u1.north)+(0,.7)$) {$\hat{w}_{12},\hat{w}_{13}$};
\draw[->] ($(u1.west)-(0.5,0)$) to (u1.west);
\node at ($(u1.west)-(.9,.25)$) {$w_{21}$};
\node at ($(u1.west)-(.9,-.25)$) {$w_{31}$};

\draw[->] ($(u2.south)+(-.3,0)$) to ($(u2.south)+(-.7,-.5)$);
\node at ($(u2.south)+(-1.2,-.75)$) {$\hat{w}_{21},\hat{w}_{23}$};
\draw[->] ($(u2.south)+(0.7,-.5)$) to ($(u2.south)+(0.3,0)$);
\node at ($(u2.south)+(1.2,-.75)$) {$w_{12},w_{32}$};

\draw[->] (u3.east) to ($(u3.east)+(.5,0)$);
\node at ($(u3.east)+(1.0,.25)$) {$\hat{w}_{31}$};
\node at ($(u3.east)+(1.0,-.25)$) {$\hat{w}_{32}$};
\draw[->] ($(u3.north)+(0,.5)$) to (u3.north);
\node at ($(u3.north)+(0,.7)$) {$w_{13},w_{23}$};
\end{tikzpicture}
\label{Fig:YCModel}
}
\caption{A $3$-way channel (3WC) and a Y-channel showing the incoming and outgoing messages at each user, and the multiplicative coefficients of the physical channels between the users.}
\end{figure*}
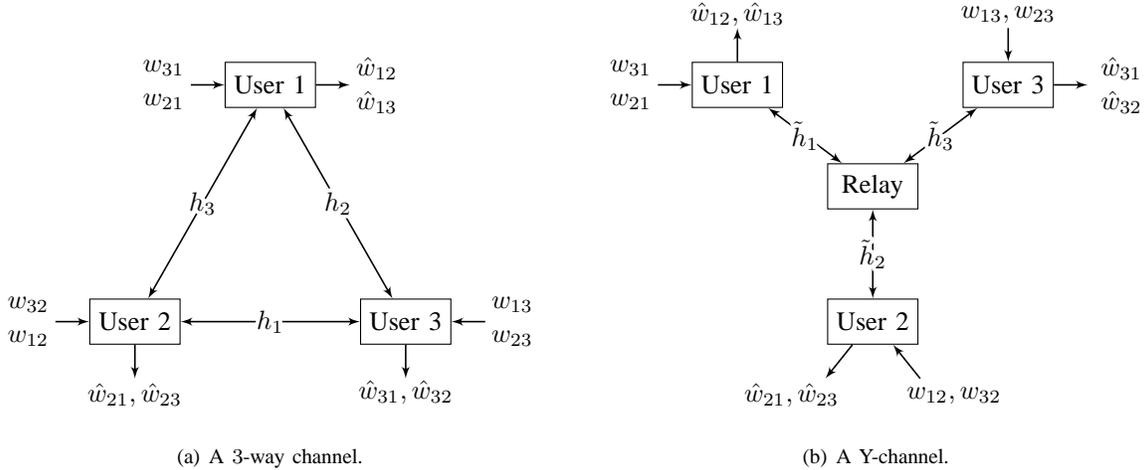

The 3WC consists of 3 FD nodes (users) communicating with each other as shown in Figure~\ref{Fig:3WCModel}. The input-output relations of this channel can be expressed as
\begin{align}
y_i(t)=h_jx_k(t)+h_kx_j(t)+z_i(t),
\end{align}
for distinct $i,j,k\in\{1,2,3\}$, where at time instant $t$, $y_i(t)$, $x_i(t)$ and $z_i(t)$ are the real-valued received signal, transmit signal, and noise at user $i$, respectively. The channel coefficient between users $j$ and $k$ is denoted by $h_i\in\mathbb{R}$ (distinct $i,j,k\in\{1,2,3\}$) and assumed to be constant during a transmission block. The noises at the receivers are independent Gaussian $\mathcal{N}(0,1)$ i.i.d. over time.

All nodes have a power constraint $P$ and have global knowledge of channel coefficients\footnote{The nodes estimate their channels before transmission starts and share the estimated values among each other.}. We assume channel reciprocity, i.e., the channel coefficient from user $i$ to user $j$ is equal to the coefficient from user $j$ to user $i$, which is a valid assumption for wireless communicating nodes sharing the same resources in an FD mode. Without loss of generality we further assume that
\begin{align}
\label{Eq:Ordering}
h_3^2\geq h_2^2\geq h_1^2.
\end{align}

User $i$ wants to communicate messages $w_{ji}$ and $w_{ki}$, uniformly distributed over the sets $\mathcal{W}_{ji}$ and $\mathcal{W}_{ki}$, with rates $R_{ji}$ and $R_{ki}$ to users $j$ and $k$, respectively. At time instant $t$, user $i$ applies an encoder 
$\mathcal{E}_{it}: \mathcal{W}_{ji}\times\mathcal{W}_{ki}\times\mathbb{R}^{t-1}\to \mathbb{R}$ to generate its transmit signal 
\begin{align}
\label{Eq:Encoder}
x_i(t)=\mathcal{E}_{it}(w_{ji},w_{ki},y_i^{(t-1)}),
\end{align} 
i.e., adaptive encoding can be used in general. The encoding functions are revealed to all users before the start of the transmission. After $n$ channel uses, user $i$ uses a decoder $\mathcal{D}_{i}: \mathcal{W}_{ji}\times\mathcal{W}_{ki}\times\mathbb{R}^{n}\to \mathcal{W}_{ij}\times\mathcal{W}_{ik}$ to decode 
\begin{align}
\label{Eq:Decoder}
(\hat{w}_{ij},\hat{w}_{ik})=\mathcal{D}_{i}(w_{ji},w_{ki},y_i^{(n)}).
\end{align}

The overall process of encoding, transmission and decoding induces an error probability $\mathbb{P}_{e,n}$ (probability of $\mathbf{w}\neq \hat{\mathbf{w}}$), where $\mathbf{w}=(w_{21},w_{31},w_{12},w_{32},w_{13},w_{23})$ and $\hat{\mathbf{w}}$ is defined similarly. A rate tuple $\mathbf{R}=(R_{21},R_{31},R_{12},R_{32},R_{13},R_{23})$ is said to be achievable if transmission at those rates can be accomplished with $\mathbb{P}_{e,n}\to0$ as $n\to\infty$. The set of all achievable rate tuples is the capacity region of the 3WC and denoted $\mathcal{C}$. Note that $\mathcal{C}$ depends on the signal-to-noise ratios ($\snr$) of the 3WC defined as $\Gamma_i=h_i^2P$, $i\in\{1,2,3\}$. We will refer to the 3WC as a 3WC$(\Gamma_1,\Gamma_2,\Gamma_3)$ channel and its capacity as $\mathcal{C}(\Gamma_1,\Gamma_2,\Gamma_3)$, which is the main focus of this work. We aim to find an approximation of $\mathcal{C}(\Gamma_1,\Gamma_2,\Gamma_3)$ within a constant gap an any $\snr$. To achieve this goal, we use the Y-channel~\cite{ChaabanSezgin_YC_Reg} as an intermediate step. The system model of the Y-channel is thus introduced next.

\subsection{System Model: Y-channel}
\label{sec:sysmod-y}
In the Y-channel, 3 FD nodes exchange information with each other in all directions via a relay node as shown in Figure \ref{Fig:YCModel}. 

The input-output relations of the Y-channel are expressed as\footnote{Throughout the paper, we will use the tilde notation to discern the Y-channel variables from the 3WC variables.}
\begin{align}
\tilde{y}_r(t)&=\tilde{h}_1\tilde{x}_1(t)+\tilde{h}_2\tilde{x}_2(t)
+\tilde{h}_3\tilde{x}_3(t)+\tilde{z}_r(t),\\
\tilde{y}_i(t)&=\tilde{h}_i\tilde{x}_r(t)+\tilde{z}_i(t),\quad i\in\{1,2,3\},
\end{align}
where at time instant $t$, $\tilde{y}_r(t)$, $\tilde{y}_i(t)$, $\tilde{x}_r(t)$, $\tilde{x}_i(t)$, $\tilde{z}_r(t)$, and $\tilde{z}_i(t)$ are the real-valued received signal at the relay, the received signal at user $i$, the transmit signal of the relay, the transmit signal of user $i$, the noise at the relay, and the noise at user $i$, respectively. The scalar $\tilde{h}_i\in\mathbb{R}$ is the channel coefficient between user $i$ and the relay, assumed to be reciprocal and to hold the same value throughout a transmission block. The noises at the receivers are independent Gaussian $\mathcal{N}(0,1)$ i.i.d. over time. All nodes have a power constraint $P$. We further assume without loss of generality that
\begin{align}
\label{Eq:OrderingY}
\tilde{h}_1^2\geq \tilde{h}_2^2\geq \tilde{h}_3^2.
\end{align}
Note that the ordering of the squared coefficients is reversed to the one given in \eqref{Eq:Ordering} for the 3WC.

In analogy to the 3WC, user $i$ wants to communicate messages $w_{ji}\in\mathcal{W}_{ji}$ and $w_{ki}\in\mathcal{W}_{ki}$ with rates $\tilde{R}_{ji}$ and $\tilde{R}_{ki}$ to users $j$ and $k$, respectively. The encoders and the decoders at the three users are similar to those applied in the 3WC. The encoding at the relay at time $t$ is done using the function $\mathcal{E}_{rt}: \mathbb{R}^{t-1}\to \mathbb{R}$ to generate $\tilde{x}_r(t)=\mathcal{E}_{rt}(\tilde{y}_r^{(t-1)})$. The set of achievable rate tuples $\tilde{\mathbf{R}}$ is the capacity region of the Y-channel, and is denoted~$\mathcal{C}_Y$. We shall denote a Y-channel as a Y$(\widetilde{\Gamma}_1,\widetilde{\Gamma}_2,\widetilde{\Gamma}_3)$ channel and its capacity region as $\mathcal{C}_Y(\widetilde{\Gamma}_1,\widetilde{\Gamma}_2,\widetilde{\Gamma}_3)$ where $\widetilde{\Gamma}_i=\tilde{h}_i^2P$, $i\in\{1,2,3\}$.

\section{Summary of the Main Results}
\label{Sec:Summary}
The main result of the paper is a capacity region approximation for the 3WC within a constant gap, as given in the following theorem.

\begin{theorem}
\label{Thm:ApproxCap}
The capacity region $\mathcal{C}(\Gamma_1,\Gamma_2,\Gamma_3)$ of a 3WC$(\Gamma_1,\Gamma_2,\Gamma_3)$ channel is within a constant gap of the region defined by the following inequalities
\begin{align}
R_{31}+R_{32} &\leq \hat{C}(\Gamma_2),\\
R_{13}+R_{23} &\leq \hat{C}(\Gamma_2), \\
R_{12}+R_{13}+R_{32} &\leq \hat{C}(\Gamma_3), \\
R_{12}+R_{13}+R_{23} &\leq \hat{C}(\Gamma_3), \\
R_{21}+R_{23}+R_{13} &\leq \hat{C}(\Gamma_3\Gamma_2/\Gamma_1), \\
R_{21}+R_{23}+R_{31} &\leq \hat{C}(\Gamma_3), \\
R_{31}+R_{32}+R_{21} &\leq \hat{C}(\Gamma_3), \\
R_{31}+R_{32}+R_{12} &\leq \hat{C}(\Gamma_3\Gamma_2/\Gamma_1),
\end{align}
with $R_{ij}\geq0$. Furthermore, the capacity region $\mathcal{C}_Y(\widetilde{\Gamma}_1,\widetilde{\Gamma}_2,\widetilde{\Gamma}_3)$ of a Y$(\widetilde{\Gamma}_1,\widetilde{\Gamma}_2,\widetilde{\Gamma}_3)$ channel with $\widetilde{\Gamma}_1=\Gamma_3\Gamma_2/\Gamma_1$, $\widetilde{\Gamma}_2=\Gamma_3$, and $\widetilde{\Gamma}_3=\Gamma_2$ is within a constant gap of this region with $R_{ij}$ replaced with $\tilde{R}_{ij}$.
\end{theorem}

This approximation is obtained by exploiting the 3WC--Y-channel ($\Delta$--Y) transformation. Therein, we treat user 1 of the 3WC as two nodes: a relay, and a `virtual' user denoted user $\tilde{1}$. The relay and user $\tilde{1}$ are connected by a virtual channel with infinite capacity. The obtained channel resembles a Y-channel, with an extra direct channel between users 2 and 3 (Fig. \ref{Fig:DYTrans2}). We call the result an extended Y-channel to discern it from the basic Y-channel. The additional link between users 2 and 3 in the extended Y-channel causes cross-talk (interference) between users 2 and 3, in both directions. It turns out that it is possible to modify the Y-channel transmission scheme in \cite{ChaabanSezgin_YC_Reg} to eliminate this cross-talk. This can be done as long as the virtual channel between the relay and user $\tilde{1}$ has an $\snr$ not less that $\Gamma_3\Gamma_2/\Gamma_1$. This modification involves introducing interference neutralization and backward decoding. After eliminating the cross-talk and setting $\tilde{\Gamma}_1=\Gamma_3\Gamma_2/\Gamma_1$, the resulting network is a Y$(\Gamma_3\Gamma_2/\Gamma_1,\Gamma_3,\Gamma_2)$ channel (Fig. \ref{Fig:DYTrans3}) whose capacity can be achieved within a constant gap using the scheme in \cite{ChaabanSezgin_YC_Reg}. Interestingly, the capacity of this transformed Y-channel is also within a constant gap of the capacity of the original 3WC, leading to the characterization in Theorem \ref{Thm:ApproxCap} above.

Proving Theorem \ref{Thm:ApproxCap} involves two parts: (i) proving it for the 3WC and (ii) for the Y-channel. The second part of the proof can be immediately obtained from \cite[Theorem 7]{ChaabanSezgin_YC_Reg}. Thus, it remains to prove the first part. The rest of the paper is devoted for proving this result, and discussing it both for the general case and also for some special cases. We start by presenting an outer bound on the capacity region $\mathcal{C}(\Gamma_1,\Gamma_2,\Gamma_3)$.

\section{Capacity Outer Bound for the 3WC}
\label{Sec:DCUpperBounds}
Upper bounds on achievable rates in the 3WC can be derived using cut-set bounds and genie-aided bounds in general. We start by stating the cut-set bounds.

\subsection{Cut-set Bounds}
The cut-set bounds for the 3WC can be written, for distinct $i,j,k\in\{1,2,3\}$, as follows \cite{CoverThomas}
\begin{align}
n(R_{ji}+R_{ki}-\varepsilon_n)&\leq I(W_{ji},W_{ki};Y_j^{(n)},Y_k^{(n)}|W_{ij},W_{kj},W_{ik},W_{jk}) \label{Eq:appx:bound1}\\
n(R_{ij}+R_{ik}-\varepsilon_n)&\leq I(W_{ij},W_{kj},W_{ik},W_{jk};Y_i^{(n)}|W_{ji},W_{ki}) \label{Eq:appx:bound2},
\end{align}
where $\varepsilon_n\to0$ as $n\to\infty$, and $W_{ij}$ and $Y_i^{(n)}$ are the random variables corresponding to $w_{ij}$ and $y_i^{(n)}$, respectively. By using Gaussian inputs, the bounds above can be further upper bounded as follows
\begin{align}
\label{Eq:CutSetBound1}
R_{ji}+R_{ki}&\leq C(h_k^2P+h_j^2P),\\
\label{Eq:CutSetBound2}
R_{ij}+R_{ik}&\leq C((|h_k|+|h_j|)^2P).
\end{align}
Details can be found in Appendix \ref{Appendix:CutSetBounds}. These bounds can be relaxed by using relation \eqref{Eq:Ordering} given by $h_3^2\geq h_2^2\geq h_1^2$ to obtain
\begin{align}
\label{CSB1}
R_{ji}+R_{ki}&\leq \hat{C}(\max\{\Gamma_k,\Gamma_j\})+1,\\
\label{CSB2}
R_{ij}+R_{ik}&\leq \hat{C}(\max\{\Gamma_k,\Gamma_j\})+3/2.
\end{align}
Tighter upper bounds on the achievable rates can be derived by using a genie-aided bounding approach as we shall see next.

\subsection{Genie-aided Bounds}
A genie-aided bound can be developed by noting that user 1 e.g. can decode $w_{32}$ given $w_{23}$, after some `noise reduction'. Consider any transmission scheme under which each user is able to decode its desired messages reliably. User 1 can thus decode $w_{12}$ and $w_{13}$. Enhance user 1 by providing $w_{23}$ and $\bar{z}_3^{(n)}=z_3^{(n)}-\frac{h_1}{h_3}z_1^{(n)}$ as side information.\footnote{Notice $\bar{z}_3^{(n)}$ effectively reduces the noise at user 1, since now user 1 can calculate $\mathbb{E}[z_1^{(n)}|\bar{z}_3^{(n)}]$ and subtract it from $y_1^{(n)}$.} Given this side information, user 1 can generate $y_3^{(n)}$ as follows. First, $w_{23}$ is combined with the decoded $w_{13}$ in order to generate $x_3(1)$ as given in \eqref{Eq:Encoder}. Then, user 1 computes $y_3(1)=\frac{h_1}{h_3}(y_1(1)-h_2x_3(1))+h_2x_1(1)+\bar{z}_3(1)$. Consequently, knowing $x_{3}(1)$ and $\bar{z}_3(1)$, user 1 can generate $y_3(1)$. Notice that at this point, knowing $w_{13}$, $w_{23}$, and $y_3(1)$ allows user 1 to generate $x_3(2)$ as given in \eqref{Eq:Encoder}. This in turn allows user 1 to compute $y_3(2)$ similar to $y_3(1)$. User 1 can repeat this procedure until all $y_3^{(n)}$ have been generated. Finally, user 1 can use $y_3^{(n)}$ together with $w_{13}$ and $w_{23}$ to decode $w_{32}$ as given in \eqref{Eq:Decoder}. 

Therefore, we can write for the enhanced user 1
\begin{align}
n(R_{12}+R_{13}+R_{32}-\varepsilon_n)\leq I\left(W_{12},W_{13},W_{32};Y_1^{(n)},\bar{Z}_3^{(n)},W_{21},W_{31},W_{23}\right), \label{Eq:appx:genie-bound1}
\end{align}
using Fano's inequality with $\varepsilon_n\to0$ as $n\to\infty$, where $W_{ij}$ is the random variable corresponding to message $w_{ij}$, and similarly $Y_1^{(n)}$ and $\bar{Z}_3^{(n)}$. After some manipulations, this bound can be recast as
\begin{align}
\label{GAB1}
R_{12}+R_{13}+R_{32}\leq \hat{C}(\Gamma_3)+2.
\end{align}
A tighter version of this bound and its detailed derivation are given in Appendix \ref{Appendix:GenieAidedBounds}. Similarly we can write the following bounds
\begin{align}
\label{GAB2}
R_{12}+R_{13}+R_{23}&\leq \hat{C}(\Gamma_3)+2\\
\label{GAB3}
R_{21}+R_{23}+R_{31}&\leq \hat{C}(\Gamma_3)+2\\
\label{GAB4}
R_{21}+R_{23}+R_{13}&\leq \hat{C}(\Gamma_3\Gamma_2/\Gamma_1)+2\\
\label{GAB5}
R_{31}+R_{32}+R_{21}&\leq \hat{C}(\Gamma_3)+2\\
\label{GAB6}
R_{31}+R_{32}+R_{12}&\leq \hat{C}(\Gamma_3\Gamma_2/\Gamma_1)+2,
\end{align}
as also shown in Appendix \ref{Appendix:GenieAidedBounds}.

By comparing the cut-set bounds \eqref{CSB1} and \eqref{CSB2} and the genie-aided bounds \eqref{GAB1}-\eqref{GAB6}, we notice that, for sufficiently high $P$, four out of the six cut-set bounds are dominated by the genie-aided ones. The only two surviving cut-set bounds are
\begin{align}
\label{CSB1R}
R_{13}+R_{23}&\leq \hat{C}(\Gamma_2)+1\\
\label{CSB2R}
R_{31}+R_{32}&\leq \hat{C}(\Gamma_2)+3/2.
\end{align}
This leads to the following Lemma.

\begin{lemma}
\label{Lemma:OuterBound}
The capacity region $\mathcal{C}(\Gamma_1,\Gamma_2,\Gamma_3)$ of the Gaussian 3WC$(\Gamma_1,\Gamma_2,\Gamma_3)$ channel is outer bounded by $\overline{\mathcal{C}}(\Gamma_1,\Gamma_2,\Gamma_3)$ where 
\begin{align}
\overline{\mathcal{C}}(\Gamma_1,\Gamma_2,\Gamma_3)=\left\{\mathbf{R}\in\mathbb{R}_+^6\left| \ \text{\eqref{GAB1}-\eqref{CSB2R} are satisfied}
\right.\right\}.
\end{align} 
\end{lemma}

Lemma \ref{Lemma:OuterBound} proves the converse of Theorem \ref{Thm:ApproxCap} for the 3WC since the outer bound $\overline{\mathcal{C}}(\Gamma_1,\Gamma_2,\Gamma_3)$ is within a constant gap of the region given in Theorem \ref{Thm:ApproxCap}. It remains to show that this region is also achievable within a constant gap. This is proved in the next section.

\section{An Achievable Rate Region for the 3WC}
\label{Sec:DCAchievability}
The outer bound $\overline{\mathcal{C}}({\Gamma}_1,{\Gamma}_2,{\Gamma}_3)$ given in Lemma \ref{Lemma:OuterBound} bears some resemblance with the approximate capacity region of the Y-channel \cite{ChaabanSezgin_YC_Reg}. In fact, the capacity region of a Y$(\widetilde{\Gamma}_1,\widetilde{\Gamma}_2,\widetilde{\Gamma}_3)$ channel with $\widetilde{\Gamma}_1=\Gamma_3\Gamma_2/\Gamma_1$, $\widetilde{\Gamma}_2=\Gamma_3$, and $\widetilde{\Gamma}_3=\Gamma_2$ is within a constant gap of $\overline{\mathcal{C}}({\Gamma}_1,{\Gamma}_2,{\Gamma}_3)$. We make use of this observation to derive an achievable rate region for the 3WC.
 
\begin{figure}[t]
\centering
\begin{tikzpicture}[semithick]
\node at (2,.75) [rectangle, fill=gray!30, draw, dashed, thin, minimum width=5.8cm, minimum height=2.65cm, rotate=0] {};

\node (u1) at (2,3.5) [rectangle, draw, thin, fill=white, minimum width=1.2cm, minimum height=.6cm, rotate=0] {User $\tilde{1}$};
\node (r) at (2,1.5) [rectangle, draw, thin, fill=white, minimum width=1.2cm, minimum height=.6cm, rotate=0] {User 1};
\node (u2) at (0,0) [rectangle, draw, thin, fill=white, minimum width=1.2cm, minimum height=.6cm, rotate=0] {User 2};
\node (u3) at (4,0) [rectangle, draw, thin, fill=white, minimum width=1.2cm, minimum height=.6cm, rotate=0] {User 3};

\draw[<->] (u1) to node[fill=white,inner sep=1pt] {\contour{white}{$\frac{h_3h_2}{h_1}$}} (r);
\draw[<->] (u2) to node[] {\contour{gray!30}{$h_3$}}  (r);
\draw[<->] (u3) to node[] {\contour{gray!30}{$h_2$}}  (r);
\draw[<->,dotted] (u3) to node[] {\contour{gray!30}{$h_1$}}  (u2);

\node at (0,1.7) {3WC};

\node at (2,1.6) [rectangle, draw, dashed, thin, minimum width=6.2cm, minimum height=4.8cm, rotate=0] {};
\node[text width=1.5cm] at (-.2,3.4) {Extended\\ \vspace{-2.5mm} Y-channel};

\node (e1) at (2,1.5) [ellipse, draw, dotted, thin, minimum width=2.0cm, minimum height=0.9cm, rotate=0] {};
\draw[->, dotted] (e1.north east) to ($(e1.north east)+(.5,.5)$);
\node at ($(e1.north east)+(1.3,.7)$) [text width=2cm] {\footnotesize{Y-channel relay}};
\end{tikzpicture}
\caption{The 3WC transformed to an extended Y-channel with user 1 acting as a relay connected to a virtual user $\tilde{1}$, and with an additional direct channel between users 2 and 3.}
\label{Fig:DeltaYC}
\end{figure}
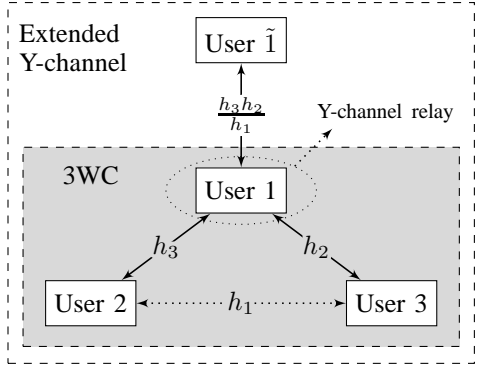

In the achievability scheme, we treat user 1 (the stronger user) as two nodes by duplicating it. We denote the duplicate as user $\tilde{1}$, which is connected to user $1$ with an infinite-capacity channel. We degrade this infinite capacity channel to a Gaussian one with $\snr$ of $\tilde{\Gamma}_1=\tilde{h}_1^2P<\infty$. The received signal at user $1$ is redefined as 
$$y_1(t)=h_2x_3(t)+h_3x_2(t)+\tilde{h}_1x_{\tilde{1}}(t)+z_1(t),$$ and we write the received signal at user $\tilde{1}$ as $$y_{\tilde{1}}(t)=\tilde{h}_1x_1(t)+z_{\tilde{1}}.$$ We choose $z_{\tilde{1}}$ as an i.i.d $\mathcal{N}(0,1)$ random variable, and $\tilde{h}_1=\frac{h_3h_2}{h_1}$. This leads to the extended Y-channel shown in Fig. \ref{Fig:DeltaYC}. 

Clearly, an achievable rate over this extended Y-channel is also achievable in the original 3WC. In what follows, we show that the achievable rate region of the basic Y-channel \cite{ChaabanSezgin_YC_Reg} (Fig. \ref{Fig:DeltaYC} with $h_1=0$) is achievable in this extended Y-channel within a constant, and hence also achievable in the 3WC within a constant. In particular, we will prove the following result.

\begin{theorem}
\label{Thm:DeltaYTransformation}
The capacity region of the 3WC given by $\mathcal{C}({\Gamma}_1,{\Gamma}_2,{\Gamma}_3)$ is within a constant gap of the capacity region of a Y-channel $\mathcal{C}_Y(\widetilde{\Gamma}_1,\widetilde{\Gamma}_2,\widetilde{\Gamma}_3)$ where $\widetilde{\Gamma}_1=\Gamma_3\Gamma_2/\Gamma_1$, $\widetilde{\Gamma}_2=\Gamma_3$, and $\widetilde{\Gamma}_3=\Gamma_2$.
\end{theorem}

To prove this, we start by discussing the simple case where $h_1=0$.

\subsection{Case $h_1=0$}
\label{h1equal0}
If $h_1=0$, then the extended Y-channel reduces to a basic Y-channel. This Y-channel has an achievable rate region given by \cite{ChaabanSezgin_YC_Reg}
\begin{align}
\label{BYC_C1}
R_{13}+R_{23}&\leq \hat{C}(\Gamma_2)-3\\
\label{BYC_C2}
R_{31}+R_{32}&\leq \hat{C}(\Gamma_2)-3\\
\label{BYC_G1}
R_{12}+R_{13}+R_{32}&\leq \hat{C}(\Gamma_3)-9/2\\
\label{BYC_G2}
R_{12}+R_{13}+R_{23}&\leq \hat{C}(\Gamma_3)-9/2\\
\label{BYC_G3}
R_{21}+R_{23}+R_{31}&\leq \hat{C}(\Gamma_3)-9/2\\
\label{BYC_G4}
R_{21}+R_{23}+R_{13}&\leq \hat{C}(\Gamma_3\Gamma_2/\Gamma_1)-9/2\\
\label{BYC_G5}
R_{31}+R_{32}+R_{21}&\leq \hat{C}(\Gamma_3)-9/2\\
\label{BYC_G6}
R_{31}+R_{32}+R_{12}&\leq \hat{C}(\Gamma_3\Gamma_2/\Gamma_1)-9/2,
\end{align}

Since this achievable rate region is within a constant gap of the outer bound $\overline{\mathcal{C}}$, then it characterizes the capacity region of the 3WC within a constant gap. Now we consider the general case $h_1\neq 0$. This proves Theorem \ref{Thm:DeltaYTransformation}, and hence also Theorem \ref{Thm:ApproxCap} for this special case. The general case is more involved due to the direct channel between users 2 and 3.

\subsection{General Case $h_1 \neq 0$}
Applying the basic Y-channel scheme on the extended Y-channel with $h_1\neq 0$ leads interference between users 2 and 3. However, this interference can be resolved using neutralization and backward decoding. To prove this, we resort to the successive-channel decomposition (SCD) described in Appendix \ref{Appendix:SCD} to simplify the treatment of the problem. We decompose the extended Y-channel using this SCD, then we modify the basic Y-channel scheme to resolve the aformentioned interference. Since this necessitates describing the basic Y-channel scheme, we revisit it next. Note that the following description is an alternative to \cite{ChaabanSezgin_YC_Reg} which does not rely on the SCD.

\subsubsection{Achievable Rate Region for the Y-channel with SCD}
\label{Sec:YCAchievability}
Consider a Y$(\widetilde{\Gamma}_1,\widetilde{\Gamma}_2,\widetilde{\Gamma}_3)$ channel. The associated uplink\footnote{We refer the transmission from the users to the relay as uplink, and the opposite direction as downlink.} is a $3\times 1$ channel with $\snr$s $\widetilde{\Gamma}_1$, $\widetilde{\Gamma}_2$, and $\widetilde{\Gamma}_3$. The downlink is a $1\times 3$ channel with $\snr$s $\widetilde{\Gamma}_1$, $\widetilde{\Gamma}_2$, and $\widetilde{\Gamma}_3$. 

Using SCD (Appendix \ref{Appendix:SCD}), the uplink is decomposed into a set of $\widetilde{N}_1\in\mathbb{N}\setminus\{0\}$ successive sub-channels, where user $i$ has access to sub-channels $1,\cdots,\widetilde{N}_i$, with $\widetilde{N}_i=\left\lfloor\frac{\log(\widetilde{\Gamma}_i)}{\log(\gamma)}\right\rfloor$ for $k=2,3$ where $\gamma^{\widetilde{N}_1}=\widetilde{\Gamma}_1$.\footnote{$\gamma$ is chosen so that $\widetilde{N}_2,\widetilde{N}_3>0$.} Similarly, the downlink is decomposed into a set of $\widetilde{N}_1$ successive sub-channels, where user $i$ has access to sub-channels $\widetilde{N}_1-\widetilde{N}_i+1,\cdots,\widetilde{N}_1$. 

An allocation of these sub-channels over the users leads to an achievable rate region for the Y-channel. The allocation used for the linear-deterministic Y-channel in \cite{ChaabanSezgin_YC_Reg} can be adopted. Denote by $\tilde{r}_{ij}$ the number of sub-channels required by user $j\in\{1,2,3\}$ to send $w_{ij}$ to user $i\neq j$. An achievable sub-channel allocation satisfies
\begin{align}
\label{SCAYC1}
\tilde{r}_{31}+\tilde{r}_{32} &\leq \widetilde{N}_3, \\
\tilde{r}_{13}+\tilde{r}_{23} &\leq \widetilde{N}_3, \\
\tilde{r}_{12}+\tilde{r}_{13}+\tilde{r}_{32} &\leq \widetilde{N}_2, \\
\tilde{r}_{12}+\tilde{r}_{13}+\tilde{r}_{23} &\leq \widetilde{N}_2, \\
\tilde{r}_{21}+\tilde{r}_{23}+\tilde{r}_{13} &\leq \widetilde{N}_1, \\
\tilde{r}_{21}+\tilde{r}_{23}+\tilde{r}_{31} &\leq \widetilde{N}_2, \\
\tilde{r}_{31}+\tilde{r}_{32}+\tilde{r}_{21} &\leq \widetilde{N}_2, \\
\label{SCAYC8}
\tilde{r}_{31}+\tilde{r}_{32}+\tilde{r}_{12} &\leq \widetilde{N}_1,
\end{align}

In this allocation, some sub-channels are allocated exclusively to one user where decode-forward is applied. Others are shared between users for the purpose of physical-layer network-coding. Shared sub-channels are used to apply bi-directional or cyclic communication \cite{ChaabanSezgin_YC_Reg}. Briefly, in bi-directional communication, two users ($i$ and $j$) send nested-lattice codewords ($\lambda_i$ and $\lambda_j$) over a sub-channel, and the relay decodes the modulo-sum of those codewords ($S(\lambda_i,\lambda_j)$) and sends it back to the two users to extract the desired codewords. In cyclic communication, users 1 and 2 send nested-lattice codewords $\lambda_{21}$ and $\lambda_{32}$, respectively, over one sub-channel, and users 2 (again) and 3 send nested-lattice codewords $\lambda_{32}$ and $\lambda_{13}$, respectively, over another sub-channels. The relay decodes the modulo-sums $S(\lambda_{21},\lambda_{32})$ and $S(\lambda_{32},\lambda_{13})$, and sends them to the users. User 1 decodes the desired signal from $S(\lambda_{21},\lambda_{32})$ and $S(\lambda_{32},\lambda_{13})$, and users 2 and 3 decode their desired signals from $S(\lambda_{21},\lambda_{32})$ and $S(\lambda_{32},\lambda_{13})$, respectively. Alternatively, users 1 and 2 send nested-lattice codewords $\lambda_{31}$ and $\lambda_{12}$, respectively, over one sub-channel, and users 1 (again) and 3 send nested-lattice codewords $\lambda_{31}$ and $\lambda_{23}$, respectively, over another sub-channels. The relay decodes the modulo-sums $S(\lambda_{31},\lambda_{12})$ and $S(\lambda_{31},\lambda_{23})$, and sends them to the users. User 2 decodes the desired signal from $S(\lambda_{31},\lambda_{12})$ and $S(\lambda_{31},\lambda_{23})$, and users 1 and 3 decode their desired signals from $S(\lambda_{31},\lambda_{12})$ and $S(\lambda_{31},\lambda_{23})$, respectively.

As long as \eqref{SCAYC1}-\eqref{SCAYC8} are satisfied, then an allocation exists which enables this communication to take place. An achievable rate region can be obtained by multiplying \eqref{SCAYC1}-\eqref{SCAYC8} by the achievable rate per sub-channel. This can be generally written as $\hat{C}(\gamma/(\kappa+\mu))$ where $\kappa$ is the number of transmitters allowed to share a sub-channel, and $\mu=0$ if the channel has one receiver and $\mu=1$ otherwise (see Sec. \ref{Sec:CommStrategies}). According to the description above, a sub-channel is used by at most two transmitters in the uplink, which has one receiver (the relay). Hence $\kappa=2$ and $\mu=0$. The downlink has one transmitter (the relay) and multiple receivers, and hence $\kappa=1$ and $\mu=1$. Thus the achievable rate per sub-channel is $\hat{C}(\gamma/2)$ in both the uplink and downlink. Note that it is required that $\gamma>2$ so that $\hat{C}(\gamma/2)>0$, which requires $\widetilde{\Gamma}_1>2^{\widetilde{N}_1}$.

Multiplying \eqref{SCAYC1}-\eqref{SCAYC8} by $\hat{C}(\gamma/2)$, and approximating $\widetilde{N}_i$ by $\frac{\log(\widetilde{\Gamma}_i)}{\log(\gamma)}$ leads to the following proposition.

\begin{proposition}
\label{Prop:YCAR}
The rate region $\underline{\mathcal{C}}_Y(\widetilde{\Gamma}_1,\widetilde{\Gamma}_2,\widetilde{\Gamma}_3,\widetilde{N}_1)$ defined by the following set of inequalities
\begin{align}
\label{ARCYC1}
\tilde{R}_{31}+\tilde{R}_{32} &\leq \hat{C}(\widetilde{\Gamma}_3)-\widetilde{N}_3/2,\\
\tilde{R}_{13}+\tilde{R}_{23} &\leq \hat{C}(\widetilde{\Gamma}_3)-\widetilde{N}_3/2, \\
\tilde{R}_{12}+\tilde{R}_{13}+\tilde{R}_{32} &\leq \hat{C}(\widetilde{\Gamma}_2)-{\widetilde{N}_2}/{2}, \\
\tilde{R}_{12}+\tilde{R}_{13}+\tilde{R}_{23} &\leq \hat{C}(\widetilde{\Gamma}_2)-{\widetilde{N}_2}/{2}, \\
\tilde{R}_{21}+\tilde{R}_{23}+\tilde{R}_{13} &\leq \hat{C}(\widetilde{\Gamma}_1)-{\widetilde{N}_1}/{2}, \\
\tilde{R}_{21}+\tilde{R}_{23}+\tilde{R}_{31} &\leq \hat{C}(\widetilde{\Gamma}_2)-{\widetilde{N}_2}/{2}, \\
\tilde{R}_{31}+\tilde{R}_{32}+\tilde{R}_{21} &\leq \hat{C}(\widetilde{\Gamma}_2)-{\widetilde{N}_2}/{2}, \\
\label{ARCYC8}
\tilde{R}_{31}+\tilde{R}_{32}+\tilde{R}_{12} &\leq \hat{C}(\widetilde{\Gamma}_1)-{\widetilde{N}_1}/{2}.
\end{align}
with $\tilde{R}_{ij}\geq0$ is achievable in a Y$(\widetilde{\Gamma}_1,\widetilde{\Gamma}_2,\widetilde{\Gamma}_3)$ channel with $\widetilde{\Gamma}_1\geq \widetilde{\Gamma}_2\geq \widetilde{\Gamma}_3$, where $\widetilde{N}_1,\widetilde{N}_2,\widetilde{N}_3\in\mathbb{N}\setminus\{0\}$, $\widetilde{N}_i=\left\lfloor\frac{\log(\widetilde{\Gamma}_i)}{\log(\gamma)}\right\rfloor$ and $\gamma^{\widetilde{N}_1}=\widetilde{\Gamma}_1>2^{\widetilde{N}_1}$.
\end{proposition}

The gap between $\underline{\mathcal{C}}_Y(\widetilde{\Gamma}_1,\widetilde{\Gamma}_2,\widetilde{\Gamma}_3,\widetilde{N}_1)$ and the outer bound given in \cite[Theorem 5]{ChaabanSezgin_YC_Reg} is $<\frac{\widetilde{N}_1+3}{4}$ bits per dimension. Since $\widetilde{N}_1$ must be chosen such that $\widetilde{N}_i>0$, and $\widetilde{\Gamma}_1>2^{\widetilde{N}_1}$, then $\widetilde{N}_1$ has to satisfy $\frac{\log(\widetilde{\Gamma}_1)}{\log(\widetilde{\Gamma}_3)}<\widetilde{N}_1<\log(\widetilde{\Gamma}_1)$. Such an $\widetilde{N}_1$ exists if $\widetilde{\Gamma}_3>2$. It is to be noted that the region \eqref{BYC_C1}-\eqref{BYC_G6}, while different from $\underline{\mathcal{C}}_Y(\widetilde{\Gamma}_1,\widetilde{\Gamma}_2,\widetilde{\Gamma}_3,\widetilde{N}_1)$, can be obtained from Proposition \ref{Prop:YCAR} by `sub-channel grouping' as explained in Appendix \ref{SubSec:SubChannelGrouping}.

\subsubsection{Resolving Interference in the Extended Y-channel}
\label{Sec:BackToThe3WC}
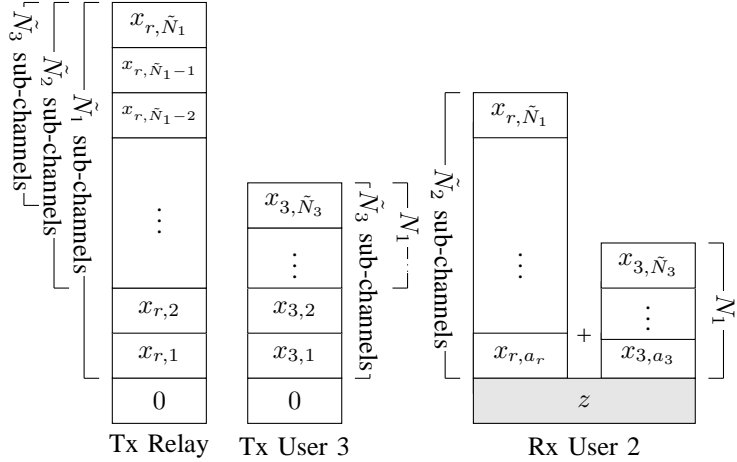
\begin{figure}
\centering
\begin{tikzpicture}

\node (x1) at (-4.8,0) [rectangle, draw, thin, fill=white, minimum width=1.25cm, minimum height=.6cm, rotate=0] {$x_{r,1}$};
\node (x2) at (-4.8,.6) [rectangle, draw, thin, fill=white, minimum width=1.25cm, minimum height=.6cm, rotate=0] {$x_{r,2}$};
\node (dots) at (-4.8,1.9) [rectangle, draw, thin, fill=white, minimum width=1.25cm, minimum height=2cm, rotate=0] {$\vdots$};
\node (xNm2) at (-4.8,3.2) [rectangle, draw, thin, fill=white, minimum width=1.25cm, minimum height=.6cm, rotate=0] {\scriptsize{$x_{r,\tilde{N}_1-2}$}};
\node (xNm1) at (-4.8,3.8) [rectangle, draw, thin, fill=white, minimum width=1.25cm, minimum height=.6cm, rotate=0] {\scriptsize{$x_{r,\tilde{N}_1-1}$}};
\node (xN) at (-4.8,4.4) [rectangle, draw, thin, fill=white, minimum width=1.25cm, minimum height=.6cm, rotate=0] {$x_{r,\tilde{N}_1}$};
\node at (-4.8,-.6) [rectangle, draw, thin, fill=white, minimum width=1.25cm, minimum height=.6cm, rotate=0] {$0$};
\draw (-5.6,4.7) to (-5.8,4.7);
\draw (-5.6,-.3) to (-5.8,-.3);
\draw (-5.8,-.3) to (-5.8,4.7);
\node at (-5.8,2.3) [fill=white,inner sep=1pt, rotate=-90] {$\tilde{N}_1$ sub-channels};

\draw (-6.0,4.7) to (-6.2,4.7);
\draw (-6.0,0.9) to (-6.2,0.9);
\draw (-6.2,0.9) to (-6.2,4.7);
\node at (-6.2,2.8) [fill=white,inner sep=1pt, rotate=-90] {$\tilde{N}_2$ sub-channels};

\draw (-6.4,4.7) to (-6.6,4.7);
\draw (-6.4,2.0) to (-6.6,2.0);
\draw (-6.6,2.0) to (-6.6,4.7);
\node at (-6.6,3.4) [fill=white,inner sep=1pt, rotate=-90] {$\tilde{N}_3$ sub-channels};

\node at (-4.8,-1.2) {Tx Relay};

\node (x1) at (-3.0,0) [rectangle, draw, thin, fill=white, minimum width=1.25cm, minimum height=.6cm, rotate=0] {$x_{3,1}$};
\node (x2) at (-3.0,.6) [rectangle, draw, thin, fill=white, minimum width=1.25cm, minimum height=.6cm, rotate=0] {$x_{3,2}$};
\node (dots) at (-3.0,1.3) [rectangle, draw, thin, fill=white, minimum width=1.25cm, minimum height=.8cm, rotate=0] {$\vdots$};
\node (xNm2) at (-3.0,2.0) [rectangle, draw, thin, fill=white, minimum width=1.25cm, minimum height=.6cm, rotate=0] {$x_{3,\tilde{N}_3}$};
\node at (-3.0,-.6) [rectangle, draw, thin, fill=white, minimum width=1.25cm, minimum height=.6cm, rotate=0] {$0$};
\draw (-2.2,2.3) to (-2.0,2.3);
\draw (-2.2,-.3) to (-2.0,-.3);
\draw (-2.0,-.3) to (-2.0,2.3);
\node at (-2.0,1.0) [fill=white,inner sep=1pt, rotate=-90] {$\tilde{N}_3$ sub-channels};

\draw (-1.7,2.3) to (-1.5,2.3);
\draw (-1.7,0.9) to (-1.5,0.9);
\draw (-1.5,1.3) to (-1.5,2.3);
\draw[dotted] (-1.5,1.3) to (-1.5,1.1);
\draw (-1.5,1.1) to (-1.5,.9);
\node at (-1.5,1.7) [fill=white,inner sep=1pt, rotate=-90] {$N_1$};

\node at (-3.0,-1.2) {Tx User 3};

\node (dots) at (0,1.3) [rectangle, draw, thin, fill=white, minimum width=1.25cm, minimum height=3.6cm, rotate=0] {$\vdots$};
\node (x1) at (0,0) [rectangle, draw, thin, fill=white, minimum width=1.25cm, minimum height=.6cm, rotate=0] {$x_{r,a_r}$};
\node (xNm2) at (0,3.2) [rectangle, draw, thin, fill=white, minimum width=1.25cm, minimum height=.6cm, rotate=0] {$x_{r,\tilde{N}_1}$};
\draw (-.8,3.5) to (-1.0,3.5);
\draw (-.8,-.3)  to (-1.0,-.3);
\draw (-1.0,-.3)  to (-1.0,3.5);
\node at (-1.0,1.3) [fill=white,inner sep=1pt, rotate=-90] {$\tilde{N}_2$ sub-channels};

\node (x1) at (1.7,0) [rectangle, draw, thin, fill=white, minimum width=1.25cm, minimum height=.6cm, rotate=0] {$x_{3,a_3}$};
\node (xd) at (1.7,.6) [rectangle, draw, thin, fill=white, minimum width=1.25cm, minimum height=.6cm, rotate=0] {$\vdots$};
\node (x2) at (1.7,1.2) [rectangle, draw, thin, fill=white, minimum width=1.25cm, minimum height=.6cm, rotate=0] {$x_{3,\tilde{N}_3}$};
\node at (.85,0.27) [] {+};
\draw (2.5,1.5) to (2.7,1.5);
\draw (2.5,-.3) to (2.7,-.3);
\draw (2.7,-.3) to (2.7,1.5);
\node at (2.7,0.6) [fill=white,inner sep=1pt, rotate=-90] {$N_1$};

\node at (0.85,-.6) [rectangle, draw, thin, fill=gray!20, minimum width=2.95cm, minimum height=.6cm, rotate=0] {$z$};
\node at (0.85,-1.2) {Rx User 2};
\end{tikzpicture}

\caption{In the extended Y-channel, the received signal at user 2 is a superposition of the relay signal, the transmit signal of user 3, and noise. Due to noise, user 2 only observes $\widetilde{N}_2$ sub-channels from the relay, i.e., $a_r=\widetilde{N}_1-(\widetilde{N}_2-1)$. On the other hand, $N_1$ sub-channels at user 2 are corrupted by the transmit signal of user 3, and thus $a_3=\widetilde{N}_3-(N_1-1)$.}
\label{Fig:ExtendedYChannelInterference}
\end{figure}

Now, we are ready to modify the basic Y-channel scheme to the extended Y-channel, by accounting for interference between users 2 and 3. Applying the SCD to the extended Y-channel leads to a similar set of sub-channels as the basic Y-channel, except that some sub-channels have a direct link between users 2 and 3. Since this interference has power $\Gamma_1$, it will corrupt the lowest $N_1=\left\lfloor\frac{\log(\Gamma_1)}{\log(\gamma)}\right\rfloor$ sub-channels at users 2 and 3 (see Fig. \ref{Fig:ExtendedYChannelInterference}). Fortunately, this interference can be nicely taken care of as we show next.

The aforementioned interference can be classified into three classes:
\begin{itemize}
 \item [(a)] Interference at user $i\in\{2,3\}$ is a dedicated signal from user $j\in\{2,3\}\setminus\{i\}$ to user $i$, that is to be forwarded by the relay (user 1) to user $i$ in the next transmission.
 \item [(b)] Interference at user 3 is a dedicated signal from user 2 to user $\tilde{1}$.
 \item [(c)] Interference at user 2 is a dedicated signal from user 3 to user $\tilde{1}$.
\end{itemize}
Next, we describe how to deal with this interference while maintaining the general structure of the basic Y-channel scheme. We start with interference of class (a).

\paragraph{Class (a) Interference}
Interference of class (a) can be canceled using backward decoding. The basic Y-channel transmission scheme is applied for $B+1$ blocks of length $n$ symbols each, where the Y-channel users are active in the first $B$ blocks, and the relay (user 1) is active in the last $B$ blocks. Decoding at the users is postponed until the end of the last block $B+1$. This is the only difference with the basic Y-channel scheme --the latter does not require backward decoding. In block $B+1$, the users are silent, and therefore, there is no interference between users 2 and 3 in this transmission block. Hence, users 2 and 3 can decode their dedicated signals corresponding to block $B+1$ from the relay signal. The users then proceed to decode the signals of block~$B$, where there is interference between users 2 and 3. However, this interference consists of desired signals (class (a)) that have been decoded by users 2 and 3 in the block $B+1$. Thus, this interference can be canceled, rendering the received signals of users 2 and 3 in block $B$ free of class (a) interference. This cancellation is performed analogously for all blocks $B-1,B-2,\cdots,1$, and class (a) interference is resolved.

\paragraph{Class (b) Interference}
\label{SubSec:ClassBInterferece}
To cancel interference of class (b), we apply the following interference neutralization scheme. User~2 \emph{pre-transmits} the interference signal one transmission block in advance as follows. Consider sub-channel $\ell\in\{1,\cdots,N_1\}$ at user 3 in downlink block~$b$, and assume that user 3 receives $\tilde{y}_{3,\ell}^{(n)}(b)=\tilde{x}_{r,\ell}^{(n)}(b)+\tilde{x}_{2,\ell}^{(n)}(b)+\tilde{z}_{3,\ell}^{(n)}(b)$ over this sub-channel, where $z_{3,\ell}^{(n)}(b)$ is the sum of noise and all received signals at sub-channels $1,\cdots,\ell-1$ at user 3. The signals $\tilde{x}_{r,\ell}^{(n)}(b)$ and $\tilde{x}_{2,\ell}^{(n)}(b)$ are nested-lattice coded as
\begin{align}
\tilde{x}_{r,\ell}^{(n)}(b)&=(\lambda_{r,\ell}(b)+d_{r,\ell}(b))\bmod\Lambda_c\\
\tilde{x}_{2,\ell}^{(n)}(b)&=(\lambda_{2,\ell}(b)+d_{2,\ell}(b))\bmod\Lambda_c,
\end{align}
where $\lambda_{i,\ell}(b)$ is a nested-lattice point, $d_{i,\ell}(b)$ is a random dither, and $\Lambda_c$ is a coarse lattice (see \cite{NazerGastpar}). Here, $\lambda_{r,\ell}(b)$ is a nested-lattice point decoded by the relay over sub-channel $\ell'$ in the uplink in block $b-1$ (due to causality), and can be generally written as $$\lambda_{r,\ell}(b)=(\lambda_{\tilde{1},\ell'}(b-1)+\lambda_{2,\ell'}(b-1)+\lambda_{3,\ell'}(b-1))\bmod\Lambda_c$$ where $\lambda_{i,\ell'}(b-1)=0$ if user~$i$ does not occupy this sub-channel.

To achieve interference neutralization, the relay should send $(\lambda_{r,\ell}(b)-\lambda_{2,\ell}(b))\bmod\Lambda_c$ instead of $\lambda_{r,\ell}$ in block~$b$ (see Appendix \ref{SubSubSection:Neutralization}). Thus, in uplink block $b-1$, sub-channel $\ell'$ should be altered by user 2 so that the relay decodes $(\lambda_{r,\ell}(b)-\lambda_{2,\ell}(b))\bmod\Lambda_c$ instead of $\lambda_{r,\ell}(b)$. To this end, user 2 pre-transmits $\lambda_{2,\ell}(b)$ to the relay by sending $x_{2,\ell'}^{(n)}(b-1)=(\lambda_{2,\ell'}(b-1)-\lambda_{2,\ell}(b)+d_{2,\ell'}(b-1))\bmod\Lambda_c$ instead of $x_{2,\ell'}^{(n)}(b-1)=(\lambda_{2,\ell'}(b-1)+d_{2,\ell'}(b-1))\bmod\Lambda_c$ over sub-channel $\ell'$ in uplink block $b-1$. The relay decodes $(\lambda_{r,\ell}(b)-\lambda_{2,\ell}(b))\bmod\Lambda_c$ in uplink block $b-1$, and consequently, user 3 receives
\begin{align*}
\tilde{y}_{3,\ell}^{(n)}(b)&=((\lambda_{r,\ell}(b)-\lambda_{2,\ell}(b))\bmod\Lambda_c+d_{r,\ell}(b))\bmod\Lambda_c+(\lambda_{2,\ell}(b)+d_{2,\ell}(b))\bmod\Lambda_c+\tilde{z}_{3,\ell}^{(n)}(b)\\
&=(\lambda_{r,\ell}(b)-\lambda_{2,\ell}(b)+d_{r,\ell}(b))\bmod\Lambda_c+(\lambda_{2,\ell}(b)+d_{2,\ell}(b))\bmod\Lambda_c+\tilde{z}_{3,\ell}^{(n)}(b),
\end{align*}
where the last step follows since $(a\bmod\Lambda+b)\bmod\Lambda=(a+b)\bmod\Lambda$ \cite{NazerGastpar}. This achieves interference neutralization after user 3 computes the sum $(\lambda_{r,\ell}(b)-\lambda_{2,\ell}(b)+\lambda_{2,\ell}(b))\bmod\Lambda_c=\lambda_{r,\ell}(b)$. 

This pre-transmission solves the problem of class (b) interference under two conditions. First, the pre-transmission should not disturb any other user. Second, user 2 should have access to sub-channel $\ell'$ from which $\lambda_{r,\ell}(b)$ is decoded by the relay in the uplink.

The first condition can be easily checked. User $\tilde{1}$ is not disturbed by this pre-transmission of $\lambda_{2,\ell}(b)$ in block $b-1$ since $\lambda_{2,\ell}(b)$ is desired by user $\tilde{1}$ (by definition of class (b) interference), and hence can be removed by backward decoding. Moreover, the pre-transmission of $\lambda_{2,\ell}(b)$ clearly does not disturb user 2 since $\lambda_{2,\ell}(b)$ originates from user 2 itself. This pre-transmission is only received at user~3 if it is sent over the top-most $N_1$ sub-channels at user 2 in the uplink. Here we distinguish between two cases: this pre-transmission is either received on sub-channel $\ell$ at user 3, or on some other sub-channel. In the latter case, this pre-transmission is combined with class (b) interference and neutralized as described above. The former case where this pre-transmission by user 2 on sub-channel $\ell'$ in the uplink interferes with sub-channel $\ell$ at user 3 in the downlink, is not possible by construction of the Y-channel scheme. Namely, this case means that the relay signal on sub-channel $\ell$ in the downlink is desired by user 3 and contains a signal from user 2 to user $\tilde{1}$ at the same time, which is not possible, see Sec. \ref{Sec:YCAchievability} (cf. \cite[Sec. VI.C.1]{ChaabanSezgin_YC_Reg}).

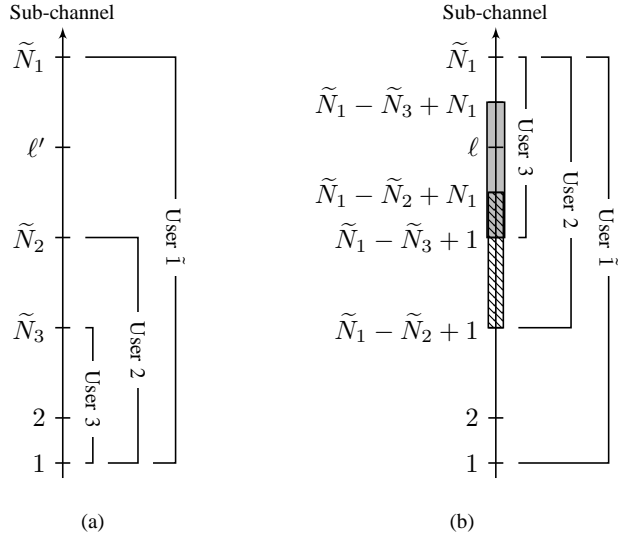
\begin{figure}
\centering
\subfigure[]{
\begin{tikzpicture}[semithick]

\draw[->] (0,-.2) to (0,5.8);
\draw[] (-.1,0) to (.1,0);
\draw[] (-.1,.6) to (.1,.6);
\draw[] (-.1,1.8) to (.1,1.8);
\draw[] (-.1,3) to (.1,3);
\draw[] (-.1,4.2) to (.1,4.2);
\draw[] (-.1,5.4) to (.1,5.4);
\node at (-.1,0) [left] {$1$};
\node at (-.1,.6) [left] {$2$};
\node at (-.1,1.8) [left] {$\widetilde{N}_3$};
\node at (-.1,3) [left] {$\widetilde{N}_2$};
\node at (-.1,4.2) [left] {$\ell'$};
\node at (-.1,5.4) [left] {$\widetilde{N}_1$};

\node at (0,6.0) [] {\footnotesize{Sub-channel}};

\draw (.3,0) -- (0.4,0) -- (.4,1.8) -- (.3,1.8);
\node at (.4,.85) [fill=white, rotate=-90] {\footnotesize User 3};
\draw (.6,0) -- (1,0) -- (1,3) -- (.3,3);
\node at (1,1.5) [fill=white, rotate=-90] {\footnotesize User 2};
\draw (1.2,0) -- (1.5,0) -- (1.5,5.4) -- (.3,5.4);
\node at (1.5,3) [fill=white, rotate=-90] {\footnotesize User $\tilde{1}$};

\end{tikzpicture}
\label{Fig:InterferenceFreeLevelsa}
}
\hspace{1cm} 
\subfigure[]{
\begin{tikzpicture}[semithick]

\node at (0,3.9) [rectangle, draw, black, minimum width= .2cm, minimum height= 1.8cm, fill=gray!50] {};


\draw[pattern=north west lines, pattern color=black] (-.1cm,1.8) rectangle (.1cm,3.6);

\draw[->] (0,-.2) to (0,5.8);
\draw[] (-.1,0) to (.1,0);
\draw[] (-.1,.6) to (.1,.6);
\draw[] (-.1,1.8) to (.1,1.8);
\draw[] (-.1,3) to (.1,3);
\draw[] (-.1,3.6) to (.1,3.6);
\draw[] (-.1,4.2) to (.1,4.2);
\draw[] (-.1,4.8) to (.1,4.8);
\draw[] (-.1,5.4) to (.1,5.4);
\node at (-.1,0) [left] {$1$};
\node at (-.1,.6) [left] {$2$};
\node at (-.1,1.8) [left] {$\widetilde{N}_1-\widetilde{N}_2+1$};
\node at (-.1,3) [left] {$\widetilde{N}_1-\widetilde{N}_3+1$};
\node at (-.1,3.6) [left] {$\widetilde{N}_1-\widetilde{N}_2+N_1$};
\node at (-.1,4.2) [left] {$\ell$};
\node at (-.1,4.8) [left] {$\widetilde{N}_1-\widetilde{N}_3+N_1$};
\node at (-.1,5.4) [left] {$\widetilde{N}_1$};

\node at (0,6.0) [] {\footnotesize{Sub-channel}};

\draw (.3,3) -- (0.4,3) -- (.4,5.4) -- (.3,5.4);
\node at (.4,4.2) [fill=white, rotate=-90] {\footnotesize User 3};
\draw (.3,1.8) -- (1,1.8) -- (1,5.4) -- (.6,5.4);
\node at (1,3.5) [fill=white, rotate=-90] {\footnotesize User 2};
\draw (0.3,0) -- (1.5,0) -- (1.5,5.4) -- (1.2,5.4);
\node at (1.5,3) [fill=white, rotate=-90] {\footnotesize User $\tilde{1}$};

\end{tikzpicture}
\label{Fig:InterferenceFreeLevelsb}
}
\caption{Sub-channels accessible by users $\tilde{1}$, 2, and 3, in the uplink (a) and downlink (b) of an extended Y-channel. The shaded and dashed sub-channels in (b) are the sub-channels at users 3 and 2 which receive interference from users 2 and 3, respectively.}
\label{Fig:InterferenceFreeLevels}
\end{figure}

The second condition needs further consideration. We need to make sure that user 2 has access to sub-channel $\ell'$ from which $\lambda_{r,\ell}$ is decoded by the relay in the uplink, i.e., $\ell'\in\{1,\cdots,\widetilde{N}_2\}$ where $\widetilde{N}_2=N_3$. The value of $\ell'$ depends on the content of $\lambda_{r,\ell}$. Recall that $\lambda_{r,\ell}$ is desired at user 3. Section \ref{Sec:YCAchievability} shows that all signals desired by user 3 are accessible by user 2 in the uplink, except the uni-directional signal from user $\tilde{1}$ to user 3. This signal, which we will refer to as $\lambda_{3\tilde{1}}^u$, might be received on sub-channels $\widetilde{N}_2+1,\cdots,\widetilde{N}_1$ in the uplink since it originates from user $\tilde{1}$ (Fig. \ref{Fig:InterferenceFreeLevelsa}). In this case, user 2 cannot alter this signal for the purpose of interference neutralization. Therefore, we need to avoid this scenario. We first note that the number of sub-channels in the problematic range $\widetilde{N}_2+1,\cdots,\widetilde{N}_1$ is $\widetilde{N}_{1}-\widetilde{N}_{2}$. On the other hand, the number of sub-channels at user 3 which do not receive any interference from user 2 is $\widetilde{N}_{3}-N_1$ (Fig. \ref{Fig:InterferenceFreeLevelsb}). By the 3WC--Y-channel transformation given in Theorem \ref{Thm:DeltaYTransformation}, we have $\widetilde{N}_{1}-\widetilde{N}_{2}=\widetilde{N}_{3}-N_1$. Therefore, we can avoid the aforementioned problematic scenario by exploiting this interesting equality. Namely, the relay of the extended Y-channel (user 1) forwards $\lambda_{3\tilde{1}}^u$ over the non-interfered downlink sub-channels at user 3 by choosing $\ell\in\{\widetilde{N}_{1}-\widetilde{N}_{3}+N_1+1,\cdots,\widetilde{N}_{1}\}$. By pursuing such an approach, the impact of class (b) interference is completely eliminated.

\paragraph{Class (c) Interference}
Class (c) interference is similar to class (b) interference. To compensate it, we apply interference neutralization again, i.e., user 3 pre-transmits the signal causing class (c) interference one transmission block in advance. As in Sec. \ref{SubSec:ClassBInterferece}, we have to guarantee two conditions. First that this pre-transmission does not disturb the users, and second that user 3 has access to all sub-channels on which this pre-transmission should take place.

By definition of class (c) interference, this pre-transmission does not affect user $\tilde{1}$ since the pre-transmitted signal itself is desired at user $\tilde{1}$, and can be canceled by backward decoding. Moreover, this pre-transmission does not affect user 3 since it originates from this same user. The pre-transmission disturbs user 2 only if it is received on the same sub-channel where this class (c) interference is received. However, this is not possible since there is no relay signal in the Y-channel scheme which is both desired at user 2, and contains a signal dedicated to user $\tilde{1}$ from user 3 (Sec. \ref{Sec:YCAchievability}). Thus, interference neutralization by pre-transmission does not disturb the users of the Y-channel.

It remains to show that user 3 has access to all sub-channels on which this pre-transmission should occur. User~3 can access $\widetilde{N}_3$ sub-channels in the uplink (Fig.~\ref{Fig:InterferenceFreeLevelsa}), whose signals are potentially forwarded to user 2 in the downlink during the next transmission block. However, user 3 cannot access sub-channels $\widetilde{N}_3+1,\cdots, \widetilde{N}_1$. Thus, if the interfered signal which is desired by user 2 is received on sub-channel $\ell\in\{\widetilde{N}_3+1,\cdots, \widetilde{N}_1\}$ by the relay in the uplink, then user 3 cannot perform interference neutralization. However, this problem is similarly solved as in the class (b) interference. The number of these problematic sub-channels is equal to the number of sub-channels at user 2 which do not receive any interference in the downlink (Fig.~\ref{Fig:InterferenceFreeLevelsb}). This follows since $\widetilde{N}_1-\widetilde{N}_3=\widetilde{N}_2-N_1$ due to $\widetilde{\Gamma}_1=\Gamma_3\Gamma_2/\Gamma_1$ (cf. Theorem \ref{Thm:DeltaYTransformation}). Thus, by sending all signals received on sub-channels $\{\widetilde{N}_3+1,\cdots, \widetilde{N}_1\}$ in the uplink on the interference-free sub-channels of user 2 in the downlink, we can guarantee that interference neutralization is performed successfully.

\subsubsection{Achievable Rate Region}
Since all interference can be resolved, it follows that the transmission scheme for the Y-channel can be applied in the extended Y-channel and hence also the 3WC. To express the achievable rate region by this scheme, we need to multiply the number of sub-channels by the achievable rate per sub-channel. According to the discussion in Appendix \ref{Sec:CommStrategies}, the achievable rate per sub-channel can be written as $C(\gamma/(\kappa+\mu))$, where $\kappa$ is the maximum number of users allowed per sub-channel, and $\mu=0$ if the number of receivers is 1, and $\mu=1$ otherwise. 

The uplink in the extended Y-channel resembles a $3\times 1$ channel, where three users can access the same sub-channel due to pre-transmissions, i.e., $\kappa=3$. The number of receivers in the uplink is 1, and thus $\mu=0$. On the other hand, the downlink resembles a $1\times 3$, where some sub-channels at user 2 are accessed by both the relay and user 3, and some sub-channels at user 3 are accessed by both the relay and user 2 (due to interference between users 2 and 3 through channel $h_1$). Thus, $\kappa=2$. Since the downlink has 3 receivers, $\mu=1$. Hence, the achievable rate per uplink and downlink sub-channel is $\hat{C}(\gamma/3)$ leading to the following proposition.

\begin{proposition}
\label{Prop:DCAR}
The rate region $\underline{\mathcal{C}}(\Gamma_1,\Gamma_2,\Gamma_3,N_3)$ defined by the following set of inequalities
\begin{align}
\label{ARCDC1}
R_{31}+R_{32} &\leq \hat{C}(\Gamma_2)-N_2\hat{C}(3),\\
R_{13}+R_{23} &\leq \hat{C}(\Gamma_2)-N_2\hat{C}(3), \\
R_{12}+R_{13}+R_{32} &\leq \hat{C}(\Gamma_3)-N_3\hat{C}(3), \\
R_{12}+R_{13}+R_{23} &\leq \hat{C}(\Gamma_3)-N_3\hat{C}(3), \\
R_{21}+R_{23}+R_{13} &\leq \hat{C}(\Gamma_3\Gamma_2/\Gamma_1)-(N_3+N_2-N_1)\hat{C}(3), \\
R_{21}+R_{23}+R_{31} &\leq \hat{C}(\Gamma_3)-N_3\hat{C}(3), \\
R_{31}+R_{32}+R_{21} &\leq \hat{C}(\Gamma_3)-N_3\hat{C}(3), \\
\label{ARCDC8}
R_{31}+R_{32}+R_{12} &\leq \hat{C}(\Gamma_3\Gamma_2/\Gamma_1)-(N_3+N_2-N_1)\hat{C}(3),
\end{align}
with $R_{ij}\geq 0$ is achievable in a 3WC$(\Gamma_1,\Gamma_2,\Gamma_3)$ channel with $\Gamma_3\geq \Gamma_2\geq \Gamma_1$, where $N_3,N_2,N_1\in\mathbb{N}\setminus\{0\}$, $N_i=\left\lfloor\frac{\log(\Gamma_i)}{\log(\gamma)}\right\rfloor$ and $\gamma^{N_3}=\Gamma_3>3^{N_3}$.
\end{proposition}

The statement of this proposition is obtained by replacing $\widetilde{\Gamma}_1$, $\widetilde{\Gamma}_2$, and $\widetilde{\Gamma}_3$ in $\underline{\mathcal{C}}_Y(\widetilde{\Gamma}_1,\widetilde{\Gamma}_2,\widetilde{\Gamma}_3,\widetilde{N}_1)$ (Proposition \ref{Prop:YCAR}) by $\Gamma_3\Gamma_2/\Gamma_1$, $\Gamma_3$, and $\Gamma_2$, respectively, which also leads to $\widetilde{N}_1=N_3+N_2-N_1$, $\widetilde{N}_2=N_3$, and $\widetilde{N}_3=N_2$. The factor $\hat{C}(3)$ arises since $\kappa+\mu=3$. Similar to Proposition \ref{Prop:YCAR}, $N_3$ should satisfy the following conditions $$\frac{\log(\Gamma_3)}{\log(\Gamma_1)}<N_3<\frac{\log(\Gamma_3)}{\log(3)},$$ and a valid $N_3$ exists if $\Gamma_1>3$.

The gap between $\underline{\mathcal{C}}(\Gamma_1,\Gamma_2,\Gamma_3,N_3)$ and $\overline{\mathcal{C}}(\Gamma_1,\Gamma_2,\Gamma_3)$ is a constant that depends mainly on $N_3$. Similar to the Y-channel, this gap can be reduced to a universal constant by using the sub-channel grouping idea explained in Section \ref{SubSec:SubChannelGrouping}. With Proposition \ref{Prop:DCAR}, the proof of Theorem \ref{Thm:DeltaYTransformation}, and hence also Theorem \ref{Thm:ApproxCap} is complete.

\subsection{Necessity of Adaptation}
The distinctive feature of the 3WC in comparison to the TWC is the necessity of adaptation. This can be clearly seen in the special case with $h_1=0$ in Sec. \ref{h1equal0}, where communication between nodes 2 and 3 is impossible without relaying through node 1. Relaying is a form of adaptation, since the transmit signal of node 1 depends not only on its message, but also on its received signal.

This is not only the case when $h_1=0$. According to Theorem \ref{Thm:ApproxCap}, the rates $R_{32}$ and $R_{23}$ are achievable within a constant if
\begin{subequations}
\label{RegTWR}
\begin{align}
R_{32} &\leq \hat{C}(\Gamma_2),\\
R_{23} &\leq \hat{C}(\Gamma_2),
\end{align}
\end{subequations}
and $R_{31}=R_{21}=R_{12}=R_{13}=0$. Note that the constraints above do not depend on $h_1$. Thus, the above rates are achievable within a constant regardless of the value of $h_1$. Indeed, the region \eqref{RegTWR} can be achieved using two-way relaying at node 1 for any $h_1$. Now suppose that adaptation is not allowed at node 1, and hence relaying is not allowed since it is a form of adaptation. In this case, nodes 2 and 3 can only communicate through their common two-way channel. The capacity of this channel is \cite{Han}
\begin{subequations}
\label{RegTWC}
\begin{align}
R_{32} &\leq C(\Gamma_1),\\
R_{23} &\leq C(\Gamma_1).
\end{align}
\end{subequations}
The gap (per dimension) between \eqref{RegTWR} and \eqref{RegTWC} is $\hat{C}(\Gamma_2)-C(\Gamma_1)=\frac{1}{2}\log\left(\frac{h_2^2P}{h_1^2P+1}\right)$ which can be arbitrarily large for small $h_1$. This implies the necessity of adaptation in the 3WC.

Note that the transformation of the 3WC to a Y-channel (Fig. \ref{Fig:DeltaYC}) `migrates' the nonadaptive role of user $1$ to user $\tilde{1}$, while keeping the adaptive role at user $1$ which now acts as a relay. This treatment simplifies designing the transmission scheme since no node has to perform both adaptive and nonadaptive roles simultaneously.

Thus, adaptation is necessary for achieving the capacity region of the 3WC within a constant, contrary to the TWC. Nevertheless, adaptation is not necessary for achieving the sum-capacity within a constant gap as shown in \cite{ChaabanMaierSezgin}. The sum-capacity is within a constant of $2\hat{C}(\Gamma_3)$, achievable via two-way communication between users 1 and 2. This sum-capacity approximation can also be concluded from Theorem \ref{Thm:ApproxCap}. Therefore, from a sum-capacity point of view, the TWC and the 3WC are similar. The TWC and the 3WC are also similar in terms of sum-capacity pre-log which is equal to 2.

\section{Special Cases}
\label{Sec:SpecialCases}
In this section, we show that our result agrees with existing results in literature for some special cases of the 3WC. We will discuss the multiple access channel (MAC) with cooperating/conferencing encoders and the broadcast channel (BC) with cooperating receivers.

\subsection{The MAC with Cooperation}
Cooperation between the users of a MAC is possible if the users can hear each others' transmission. From this point of view, a 2-user MAC with cooperation can be modeled as a 3WC, with a specific message exchange scenario. In what follows, we discuss the scenario where the users are connected by the stronger channel $h_3$. This is motivated by D2D communication in cellular networks, where users within close proximity can communicate with each other simultaneously while communicating with a base-station. A similar discussion can be pursued for other cases where the cooperation channel is weaker than one or both the channels to the base-station.

\subsubsection{Conferencing}
The MAC with cooperating transmitters was first studied by Willems \cite{Willems} in the so-called MAC with conferencing encoders. In this channel, users 1 and 2 want to communicate two independent messages $W_{1}$ and $W_{2}$ to a common receiver, respectively. The users are allowed to communicate with each other in a conferencing phase over $h_3$ before they communicate with the receiver in the transmission phase. The conferencing takes place over two error-free channels with capacities $C_{21}$ (from user 1 to 2) and $C_{12}$ (from user 2 to 1). Let each of the two users have a power constraint $P$, and let the common receiver's signal be given by $$y=h_2x_1+h_1x_2+z$$
where $z$ is a Gaussian noise with zero mean and unit variance. Thus, the $\snr$'s of the channels from users~1 and~2 to the receiver are $\Gamma_2$ and $\Gamma_1$, respectively. The capacity region of this channel was given in \cite{BrossLapidothWigger}~as $$\mathcal{C}_{\text{MAC,conf}}=\bigcup\mathcal{R}_{\text{MAC,conf}}(\beta_1,\beta_2),$$ where the union is over $\beta_1,\beta_2\in[0,1]$, $\mathcal{R}_{\text{MAC,conf}}(\beta_1,\beta_2)$ is the set of rate pairs $(R_1,R_2)\in\mathbb{R}_+^2$ satisfying
\begin{align}
R_1&\leq C(\beta_1\Gamma_2)+C_{21},\\
R_2&\leq C(\beta_2\Gamma_1)+C_{12},\\
R_1+R_2&\leq C(\beta_1\Gamma_2+\beta_2\Gamma_1)+C_{21}+C_{12},\\
R_1+R_2&\leq C(\Gamma_1+\Gamma_2+2\sqrt{\bar{\beta}_1\bar{\beta}_2\Gamma_1\Gamma_2}),
\end{align}
and $\bar{\beta}_1=1-\beta_1$ and $\bar{\beta}_2=1-\beta_2$. Now assume that $C_{12}=C_{21}=C(\Gamma_3)$. Furthermore, assume that $\Gamma_3\geq \Gamma_2\geq \Gamma_1$. Under these assumptions, the MAC with conferencing encoders resembles a 3WC with $R_{31}=R_1$, $R_{32}=R_2$, and $R_{12}=R_{13}=R_{21}=R_{23}=0$. In this case, it can be easily verified that the region $\mathcal{C}_{\text{MAC,conf}}$ is within a constant gap of the region defined by
\begin{align}
R_1+R_2&\leq \hat{C}(\Gamma_2).
\end{align}
But the latter region coincides with the statement of Theorem \ref{Thm:ApproxCap} for this special case. Thus, the statements of Theorem \ref{Thm:ApproxCap} and \cite{BrossLapidothWigger} agree within a constant gap, and our result characterizes the capacity region of the MAC with conferencing encoders within a constant gap. This leads to the following interesting consequence.

\begin{remark}
\label{Rem:StrongCoop}
If the conferencing channel has high capacity given by $C_{12}=C_{21}=C(\Gamma_3)$ with $\Gamma_3\geq \Gamma_2,\Gamma_1$, then conferencing can be replaced by in-band cooperation with marginal impact on the capacity region.
\end{remark}

Here, by marginal impact we mean that the impact is bounded by a constant independent of the $\snr$'s. Thus, in this case, we do not need to spend any extra time for conferencing, since cooperation can be established simultaneously while transmitting to the receiver.

\subsubsection{In-band cooperation}
The MAC with in-band cooperation (also known as the MAC with generalized feedback \cite{Willems}) differs from the MAC with conferencing in that cooperation takes place simultaneously with transmission to the receiver. Consider a Gaussian MAC with cooperation, where two users have a power constraint $P$, and where the received signals are given by 
\begin{align}
y&=h_2x_1+h_1x_2+z,\\
y_1&=h_3x_2+z_1,\\
y_2&=h_3x_2+z_2,
\end{align}
where $z$, $z_1$, and $z_2$ are independent Gaussian noises with zero mean and unit variance. The $\snr$'s of the channels from users 1 and 2 to the receiver are $\Gamma_2$ and $\Gamma_1$, respectively, and the $\snr$ of the cooperation channel is $\Gamma_3$. An achievable rate region for this scenario is given by \cite{SendonarisErkipAazhang} $$\mathcal{R}_{\text{MAC,coop}}=\bigcup\mathcal{R}_{\text{MAC,coop}}(\boldsymbol{\beta}_{1},\boldsymbol{\beta}_{2}),$$ where the union is over $\boldsymbol{\beta}_{i}=(\beta_{i1},\beta_{i2},\beta_{i3})$ satisfying $\beta_{ij}\geq 0$ and $\sum_{j=1}^3\beta_{ij}\leq 1$ for $i=1,2$ and $j=1,2,3$, and where $\mathcal{R}_{\text{MAC,coop}}(\boldsymbol{\beta}_{1},\boldsymbol{\beta}_{2})$ is the set of rate pairs $(R_1,R_2)\in\mathbb{R}_+^2$ satisfying
\begin{align}
\label{RiMACC}
R_i&\leq C\left(\frac{\beta_{i1}\Gamma_3}{1+\beta_{i2}\Gamma_3}\right)
+C\left(\beta_{i2}\Gamma_j\right),\ i,j\in\{1,2\},\ i\neq j\\
R_1+R_2&\leq C\left(\Gamma_2+\Gamma_1+2\sqrt{\beta_{13}\beta_{23}\Gamma_1\Gamma_2}\right)\\
R_1+R_2&\leq C\left(\beta_{12}\Gamma_2+\beta_{22}\Gamma_1\right)
+\sum_{i=1}^2C\left(\frac{\beta_{i1}\Gamma_3}{1+\beta_{i2}\Gamma_3}\right).
\end{align}
For $\Gamma_3\geq \Gamma_2\geq \Gamma_1$, it can be easily shown that the above region is within a constant gap of the region
\begin{align}
R_1+R_2&\leq \hat{C}(\Gamma_2).
\end{align}

The latter region coincides with the statement of Theorem \ref{Thm:ApproxCap} for this case as shown for the conferencing MAC above. Thus, the statements of Theorem \ref{Thm:ApproxCap} and the achievable rate region $\mathcal{R}_{\text{MAC,coop}}$ obtained from \cite{SendonarisErkipAazhang} agree within a constant gap. Furthermore, our result provides an outer bound on the capacity region for this case, and thus shows that the region $\mathcal{R}_{\text{MAC,coop}}$ characterizes the capacity region of the MAC with in-band cooperation within a constant gap. Note that this statement confirms Remark \ref{Rem:StrongCoop}.

\subsection{The BC with Cooperation}
Similar to the MAC with cooperation, it is possible to establish cooperation in a broadcast channel (BC) if users hear each others' transmission. A 2-user BC with cooperation is thus also a special case of a 3WC. Next, we consider a scenario where the receivers of the BC are connected by the stronger channel $h_3$, motivated by D2D communications. Other cases can be discussed similarly.

The input-output relationships of a BC with in-band cooperation can be written as
\begin{align}
y_1&=h_1x_3+h_3x_2+z_1,\\
y_2&=h_2x_3+h_3x_1+z_2,
\end{align}
where $z_1$ and $z_2$ are independent Gaussian noises with zero mean and unit variance. Each of the nodes has power~$P$. Therefore, the $\snr$'s corresponding to channels $h_1$, $h_2$, and $h_3$ are $\Gamma_1$, $\Gamma_2$, and $\Gamma_3$, respectively. An achievable rate region for this channel was given in \cite{LiangVeeravalli} as
$$\mathcal{R}_{\text{BC,coop}}=\bigcup\mathcal{R}_{\text{MAC,coop}}(\beta_2,\boldsymbol{\beta}_3),$$
where the union is over $\beta_2\in[0,1]$, and $\boldsymbol{\beta}_3=(\beta_{31},\beta_{32},\beta_{33})$ with $\beta_{3j}\geq0$ and $\sum_{j=1}^3\beta_{3j}\leq 1$, and where $\mathcal{R}_{\text{MAC,coop}}(\beta_2,\boldsymbol{\beta}_3)$ is the set of rate pairs $(R_1,R_2)\in\mathbb{R}_+^2$ satisfying
\begin{align}
R_1&\leq C\left(\beta_{31}\Gamma_2+\frac{\beta_{31}\beta_2\Gamma_1\Gamma_3}{1+\beta_2\Gamma_3+\beta_{31}(\Gamma_1+\Gamma_2)}\right)\\
R_2&\leq C\left(\frac{(\beta_{33}+\beta_{32})\Gamma_1+\Gamma_3+2\sqrt{\beta_{32}\Gamma_1\Gamma_3}}{1+\beta_{31}\Gamma_1}\right)\\
R_2&\leq C\left(\frac{\beta_{33}\Gamma_2}{1+\beta_{31}\Gamma_2+\beta_2\Gamma_3}\right).
\end{align}
The region $\mathcal{R}_{\text{BC,coop}}$ is within a constant gap of the region
\begin{align}
R_1+R_2 &\leq \hat{C}(\Gamma_2),
\end{align}
which is the approximate capacity region of the 3WC with $\Gamma_3\geq \Gamma_2\geq \Gamma_1$, and with $R_{31}=R_1$, $R_{32}=R_2$, $R_{12}=R_{13}=R_{21}=R_{23}=0$. In particular, by setting $\boldsymbol{\beta}_3=(1,0,0)$ we achieve $R_1\leq C(\Gamma_2)$, and by setting $\boldsymbol{\beta}_3=(0,0,1)$ and $\beta_2=0$ we achieve $R_2\leq C(\Gamma_2)$. Time-sharing between these two solutions is within a constant of $R_1+R_2 \leq \hat{C}(\Gamma_2)$.

Therefore, also for the BC with cooperation, the statement of Theorem \ref{Thm:ApproxCap} and the achievable rate region $\mathcal{R}_{\text{BC,coop}}$ obtained from \cite{LiangVeeravalli} agree within a constant gap in this case (cooperation channel $h_3$). Thus, $\mathcal{R}_{\text{BC,coop}}$ is within a constant gap of the capacity region of the BC with in-band cooperation.

Finally, we note that similar analysis can be applied for the relay channel and the two-way relay channel and variations thereof. The capacities of those channels are within a constant gap of $\mathcal{C}(\Gamma_1,\Gamma_2,\Gamma_3)$ in Theorem \ref{Thm:ApproxCap}.

\section{Conclusion}
\label{Sec:Conc}
We have studied the capacity region of the 3-way channel consisting of 3 users communicating in all directions, denoted 3WC. First, we derived a capacity region outer bound for the 3WC. A resemblance between the derived outer bound and the approximate capacity region of the Y-channel in \cite{ChaabanSezgin_YC_Reg} motivated us to exploit a 3WC--Y-channel ($\Delta$--Y) transformation to develop a transmission scheme for the 3WC. The transmission scheme is based on a modified version of the Y-channel scheme in \cite{ChaabanSezgin_YC_Reg} which is suited for the 3WC, where the additional ingredients are backward decoding and interference neutralization. The achievable rate region is shown to be within a constant gap of the derived outer bound, leading to the desired approximate capacity characterization. An interesting difference with the two-way channel is the necessity of adaptation in the 3 user case.

As a side contribution, we provide a systematic approach for decomposing a Gaussian channel into a set of sub-channels, which simplifies the search for achievable rate regions. This decomposition generalizes the treatments used in \cite{BreslerParekhTse_IT,BreslerParekhTse,
SridharanJafarianVishwanathJafarShamai,
ChaabanSezgin_IT_IRC,SahaBerry,
SezginAvestimehrKhajehnejadHassibi,VahidSuhAvestimehr} to a topology independent decomposition.

The interesting relation between the 3WC and the Y-channel raises several interesting questions. Does this transformation extend to larger networks (of more than 3 users) with mesh and star topologies? Likewise, can rather sophisticated networks be broken down into simpler networks by relations similar to the 3WC--Y-channel transformation? Answering these questions would greatly simplify studying the information-theoretic limits of larger networks.

\begin{appendices}

\section{Cut-set bounds for Gaussian Inputs}
\label{Appendix:CutSetBounds}
The first cut-set bound we consider is \eqref{Eq:appx:bound1}, i.e.,
\begin{align}
n(R_{ji}+R_{ki}-\varepsilon_n)&\leq I(W_{ji},W_{ki};Y_j^{(n)},Y_k^{(n)}|W_{ij},W_{kj},W_{ik},W_{jk})\\
&= h(Y_j^{(n)},Y_k^{(n)}|W_{ij},W_{kj},W_{ik},W_{jk})-h(Y_j^{(n)},Y_k^{(n)}|\mathbf{W}),
\end{align}
where $\varepsilon_n\to0$ as $n\to\infty$ and $\mathbf{W}=(W_{ij},W_{ji},W_{ik},W_{ki},W_{kj},W_{jk})$. The conditional joint differential entropy of $Y_j^{(n)}$ and $Y_k^{(n)}$ can be upper bounded as follows
\begin{align}
&\hspace{-.5cm}h(Y_j^{(n)},Y_k^{(n)}|W_{ij},W_{kj},W_{ik},W_{jk})\nonumber\\
\label{CSB1T0}
&= \sum_{t=1}^n h(Y_j(t),Y_k(t)|W_{ij},W_{kj},W_{ik},W_{jk},Y_j^{t-1},Y_k^{t-1})\\
&= \sum_{t=1}^n h(Y_j(t),Y_k(t)|W_{ij},W_{kj},W_{ik},W_{jk},Y_j^{t-1},Y_k^{t-1},X_j^t,X_{k}^t)\\
&= \sum_{t=1}^n h(h_kX_i(t)+Z_j(t),h_jX_i(t)+Z_k(t)|W_{ij},W_{kj},W_{ik},W_{jk},Y_j^{t-1},Y_k^{t-1},X_j^t,X_k^t)\\
&\leq \sum_{t=1}^n h(h_kX_i(t)+Z_j(t),h_jX_i(t)+Z_k(t))\\
\label{CSB1T1}
&\leq \frac{n}{2}\log((2\pi e)^2(1+h_k^2P+h_j^2P)),
\end{align}
where in \eqref{CSB1T0}-\eqref{CSB1T1}, we have used the chain rule, \eqref{Eq:Encoder}, $h(X|Y)={h(X-Y|Y)}$, the fact that conditioning does not increase entropy, and that the Gaussian distribution maximizes differential entropy under a covariance constraint \cite{CoverThomas}, respectively. 

The differential entropy term $h(Y_j^{(n)},Y_k^{(n)}|\mathbf{W})$ can be bounded by the entropy of the noise random variables $Z_j$ and $Z_k$ as follows
\begin{align}
\label{NoiseEntropy0}
h(Y_j^{(n)},Y_k^{(n)}|\mathbf{W})&=\sum_{t=1}^nh(Y_j(t),Y_k(t)|\mathbf{W},Y_j^{t-1},Y_k^{t-1})\\
&\geq \sum_{t=1}^nh(Y_j(t),Y_k(t)|\mathbf{W},Y_j^{t-1},Y_k^{t-1},Y_i^{t-1})\\
&= \sum_{t=1}^nh(Y_j(t),Y_k(t)|\mathbf{W},Y_j^{t-1},Y_k^{t-1},Y_i^{t-1},X_j^{t},X_k^{t},X_i^{t})\\
&= \sum_{t=1}^nh(Z_j(t),Z_k(t)|\mathbf{W},Z_j^{t-1},Z_k^{t-1},Z_i^{t-1},X_j^{t},X_k^{t},X_i^{t})\\
&= \sum_{t=1}^nh(Z_j(t),Z_k(t)|\mathbf{W},Z_j^{t-1},Z_k^{t-1},Z_i^{t-1})\\
&= \sum_{t=1}^nh(Z_j(t),Z_k(t))\\
\label{NoiseEntropy}
&=\frac{n}{2}\log((2\pi e)^2),
\end{align}
where steps \eqref{NoiseEntropy0}-\eqref{NoiseEntropy} follow by using the chain rule, the fact that conditioning does not increase entropy, \eqref{Eq:Encoder}, $h(X|Y)=h(X-Y|Y)$, the fact that $X_j^{t}$, $X_k^{t}$, and $X_i^{t}$ can be constructed from $\mathbf{W}$, $Z_j^{t-1}$, $Z_k^{t-1}$, and $Z_i^{t-1}$ (since $h_i$, $h_j$, and $h_k$ are known at all nodes), the independence of the noises and the messages, and that $Z_j$ and $Z_k$ are $\mathcal{N}(0,1)$. Combining \eqref{NoiseEntropy} and \eqref{CSB1T1}, dividing by $n$, and letting $n\to\infty$ yields the bound \eqref{Eq:CutSetBound1}.

The second cut-set bound \eqref{Eq:appx:bound2} is given by
\begin{align}
n(R_{ij}+R_{ik}-\varepsilon_n)&\leq I(W_{ij},W_{ik};Y_i^{(n)}|W_{ji},W_{jk},W_{ki},W_{kj})\\
&= h(Y_i^{(n)}|W_{ji},W_{jk},W_{ki},W_{kj})-h(Y_i^{(n)}|\mathbf{W}).
\end{align}
Similar to above, by using the chain rule, the fact that conditioning does not increase entropy, and that the Gaussian distribution is a differential entropy maximizer, we get
\begin{align}
h(Y_i^{(n)}|W_{ji},W_{jk},W_{ki},W_{kj})&\leq \frac{n}{2}\log(2\pi e(1+h_j^2P+h_k^2P+2h_jh_k\rho P))\\
\label{CSB2T1}
&\leq \frac{n}{2}\log(2\pi e(1+(|h_j|+|h_k|)^2P)),
\end{align}
where $\rho\in[-1,1]$ is the correlation between the Gaussian $X_j$ and $X_k$. Similarly, we can show that
\begin{align}
\label{NoiseEntropy2}
h(Y_i^{(n)}|\mathbf{W})\geq \frac{1}{2}\log(2\pi e).
\end{align}
Combining \eqref{NoiseEntropy2} and \eqref{CSB2T1}, dividing by $n$, and letting $n\to\infty$ yields the desired bound \eqref{Eq:CutSetBound2}.

\section{Genie-aided Bounds}
\label{Appendix:GenieAidedBounds}
We start with the bound \eqref{Eq:appx:genie-bound1}, i.e.,
\begin{align}
n(R_{12}+R_{13}+R_{32}-\varepsilon_n)\leq I(W_{12},W_{13},W_{32};Y_1^{(n)},\bar{Z}_3^{(n)},W_{21},W_{31},W_{23}),
\end{align}
where $Z_3^{(n)}-\frac{h_1}{h_3}Z_1^{(n)}$ by $\bar{Z}_3^{(n)}$, which we can rewrite as
\begin{align}
n(R_{12}+R_{13}+R_{32}-\varepsilon_n)\leq I(W_{12},W_{13},W_{32};Y_1^{(n)},\bar{Z}_3^{(n)}|W_{21},W_{31},W_{23}),
\end{align}
due to the independence of the messages. By using the definition of mutual information, we get
\begin{align}
\label{UB1T0}
n(R_{12}+R_{13}+R_{32}-\varepsilon_n)&\leq h(Y_1^{(n)},\bar{Z}_3^{(n)}|W_{21},W_{31},W_{23})-h(Y_1^{(n)},\bar{Z}_3^{(n)}|\mathbf{W}).
\end{align}
Now we proceed by bounding each of the terms above separately. First we consider the term $h(Y_1^{(n)},\bar{Z}_3^{(n)}|W_{21},W_{31},W_{23})$ which we can bound as follows
\begin{align}
h(Y_1^{(n)},\bar{Z}_3^{(n)}|W_{21},W_{31},W_{23})&\leq h(Y_1^{(n)},\bar{Z}_3^{(n)})\\
&=\sum_{t=1}^n h(Y_{1}(t),\bar{Z}_3(t)|Y_{1}^{t-1},\bar{Z}_3^{t-1})\\
\label{UB1T1}
&\leq \sum_{t=1}^n h(Y_{1}(t),\bar{Z}_3(t))\\
&= \sum_{t=1}^n \left[h(\bar{Z}_3(t))+h(Y_{1}(t)|\bar{Z}_3(t))\right].
\end{align}
Next, we consider $h(Y_1^{(n)},\bar{Z}_3^{(n)}|\mathbf{W})$ which we bound as follows
\begin{align}
h(Y_1^{(n)},\bar{Z}_3^{(n)}|\mathbf{W})&=\sum_{t=1}^n h(Y_1(t),\bar{Z}_3(t)|\mathbf{W}, Y_1^{t-1},\bar{Z}_3^{t-1})\\
&\geq \sum_{t=1}^n h(Y_1(t),\bar{Z}_3(t)|\mathbf{W}, Y_1^{t-1},\bar{Z}_3^{t-1},X_1^t,X_2^t,X_3^t)\\
&= \sum_{t=1}^n h(Z_1(t),\bar{Z}_3(t)|\mathbf{W}, Z_1^{t-1},\bar{Z}_3^{t-1},X_1^t,X_2^t,X_3^t)\\
\label{UB1T2}
&= \sum_{t=1}^n h(Z_1(t),\bar{Z}_3(t)) \\
&= \sum_{t=1}^n [h(Z_1(t))+h(\bar{Z}_3(t)|Z_1(t))]
\end{align}
where that last but one step follows since the noise at time instant $t$ is independent of all past noise samples, the messages, and the transmit signals up to time instant $t$ (only the transmit signals at times $t+1,\cdots,n$ can be dependent on the noise samples at time $t$ \eqref{Eq:Encoder}). By plugging \eqref{UB1T1} and \eqref{UB1T2} in \eqref{UB1T0} we obtain
\begin{align}
n(R_{12}+R_{13}+R_{32}-\varepsilon_n)&\leq \sum_{t=1}^n \left[h(Y_{1}(t),\bar{Z}_3(t))-h(Z_1(t),\bar{Z}_3(t))\right].
\end{align}
This can be rewritten as follows
\begin{align}
n(R_{12}+R_{13}+R_{32}-\varepsilon_n)&\leq \sum_{t=1}^n \left[h(Y_{1}(t)|\bar{Z}_3(t))-h(Z_1(t)|\bar{Z}_3(t))\right]
\end{align}
by using the chain rule. Using \cite[Lemma 6]{AnnapureddyVeeravalli}, we can rewrite this bound as
\begin{align}
n(R_{12}+R_{13}+R_{32}-\varepsilon_n)&\leq \sum_{t=1}^n \left[h(h_2X_3(t)+h_3X_2(t)+V(t))-h(V(t))\right],
\end{align}
where $V\sim\mathcal{N}(0,\frac{h_3^2}{h_1^2+h_3^2})$. Now by using the Gaussian distribution for $X_2$ and $X_3$, we can maximize this bound to obtain
\begin{align}
n(R_{12}+R_{13}+R_{32}-\varepsilon_n)&\leq nC\left(\frac{h_1^2+h_3^2}{h_3^2}(|h_2|+|h_3|)^2P\right).
\end{align}
By dividing by $n$, and letting $n\to\infty$ we obtain
\begin{align}
R_{12}+R_{13}+R_{32}&\leq C\left(\frac{h_1^2+h_3^2}{h_3^2}(|h_2|+|h_3|)^2P\right).
\end{align}
Next, we relax this bound by using $h_3^2\geq h_2^2\geq h_1^2$ (cf. \eqref{Eq:Ordering}) as follows
\begin{align}
R_{12}+R_{13}+R_{32}&\leq C\left(\frac{h_1^2+h_3^2}{h_3^2}(|h_2|+|h_3|)^2P\right)\\
&\leq C(2(|h_2|+|h_3|)^2P)\\
&\leq C(2(2|h_3|)^2P)\\
&= C(8h_3^2P)\\
&\leq \frac{1}{2}\log(h_3^2P)+2,
\end{align}
which is the desired upper bound in \eqref{GAB1}. The remaining bounds \eqref{GAB2}-\eqref{GAB6} can be derived similarly by giving each node the suitable side-information. Namely, \eqref{GAB2} can be derived by giving $W_{32}$ and $Z_2^{(n)}-\frac{h_1}{h_2}Z_1^{(n)}$ to node 1 as side information. The third and fourth bounds \eqref{GAB3} and \eqref{GAB4} can be derived by giving $(W_{13},Z_3^{(n)}-\frac{h_2}{h_3}Z_2^{(n)})$ and $(W_{31},Z_1^{(n)}-\frac{h_2}{h_1}Z_2^{(n)})$ to node 2 as side information, respectively. Finally, the side information that should be given to node 3 in order to obtain the bounds \eqref{GAB5} and \eqref{GAB6} are $(W_{12},Z_2^{(n)}-\frac{h_3}{h_2}Z_3^{(n)})$ and $(W_{21},Z_1^{(n)}-\frac{h_3}{h_1}Z_3^{(n)})$, respectively.

\section{Successive Channel Decomposition (SCD)}
\label{Appendix:SCD}

The decomposition described in this appendix is similar to approaches used in\cite{BreslerParekhTse_IT,BreslerParekhTse,
SridharanJafarianVishwanathJafarShamai,
ChaabanSezgin_IT_IRC,SahaBerry,
SezginAvestimehrKhajehnejadHassibi,VahidSuhAvestimehr}. Here, we generalize it to a topology-independent decomposition.

\subsection{Point-to-Point Channel (P2P)}
Consider a P2P channel with input $X$ satisfying $\mathbb{E}[X^2]\leq \Gamma$, output $Y=X+Z$, and noise $Z\sim\mathcal{N}(0,1)$, i.i.d. over time. We would like to decompose this channel into a set of $N$ sub-channels. To this end, we write the $\snr$ of the channel ($\Gamma$) as
\begin{align}
\label{GammaDecomposed}
\Gamma=1+\sum_{\ell=1}^Np_\ell,
\end{align}
where $p_\ell=\gamma^\ell-\gamma^{\ell-1}$ for a given $\gamma\in\mathbb{R}$ with $\gamma^N=\Gamma$.

Let the length-$n$ transmit signal $x^{(n)}=(x(1),\cdots,x(n))$ be given as $x^{(n)}=\sum_{\ell=1}^Nx_\ell^{(n)}$ where $x_\ell^{(n)}$ has power $p_\ell$ and rate $R_\ell$. This multi-level coded signal satisfies the power constraint. The receiver decodes successively, by decoding $x_\ell^{(n)}$ while treating $x_1^{(n)},\cdots,x_{\ell-1}^{(n)}$ as noise starting with $\ell=N$ and ending with $\ell=1$. At each decoding step, the decoded signals are subtracted from the received signal. The achievable rates can be written as
\begin{align}
R_\ell 
\label{Eq:P2PperSubChannelRate}
&= \hat{C}\left(\frac{\gamma^\ell}{\gamma^{\ell-1}}\right)=\frac{1}{N}\hat{C}\left(\Gamma\right).
\end{align}
We shall call the cumulative power of the first $\ell$ signals, i.e., $\gamma^\ell$, a {\it `power-level'}. To obtain the total achievable rate, we simply multiply \eqref{Eq:P2PperSubChannelRate} by $N$ to obtain $R=\sum_{\ell=1}^NR_\ell= \hat{C}(\Gamma)$. This rate is only meaningful if $\Gamma>1$, and it converges to the channel capacity as $\Gamma$ increases.

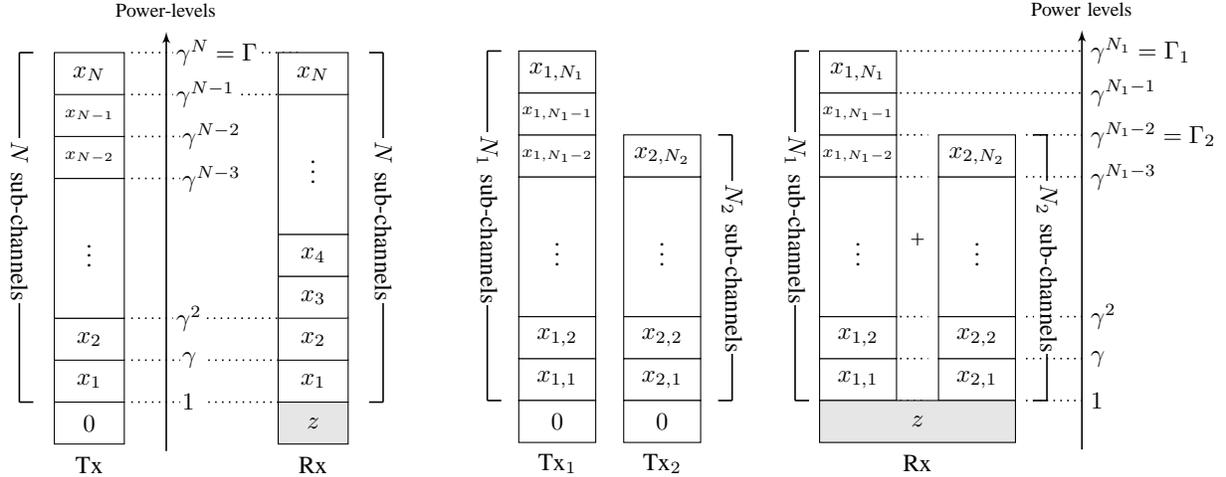
\begin{figure*}
\centering
\subfigure[A point-to-point channel as a set of successive sub-channels indicated by the $N$ power-levels $1,\gamma,\cdots,\gamma^N$.]{
\begin{tikzpicture}[semithick,scale=.93, every node/.style={scale=.93}]]
\node (x1) at (0,0) [rectangle, draw, thin, fill=white, minimum width=1.0cm, minimum height=.6cm, rotate=0] {$x_{1}$};
\node (x2) at (0,.6) [rectangle, draw, thin, fill=white, minimum width=1.0cm, minimum height=.6cm, rotate=0] {$x_{2}$};
\node (dots) at (0,1.9) [rectangle, draw, thin, fill=white, minimum width=1.00cm, minimum height=2cm, rotate=0] {$\vdots$};
\node (xNm2) at (0,3.2) [rectangle, draw, thin, fill=white, minimum width=1.00cm, minimum height=.6cm, rotate=0] {\scriptsize{$x_{N-2}$}};
\node (xNm1) at (0,3.8) [rectangle, draw, thin, fill=white, minimum width=1.0cm, minimum height=.6cm, rotate=0] {\scriptsize{$x_{N-1}$}};
\node (xN) at (0,4.4) [rectangle, draw, thin, fill=white, minimum width=1.0cm, minimum height=.6cm, rotate=0] {$x_{N}$};
\node at (0,-1.2) {Tx};
\node at (0,-.6) [rectangle, draw, thin, fill=white, minimum width=1.0cm, minimum height=.6cm, rotate=0] {$0$};
\draw (-.8,4.7) to (-1.0,4.7);
\draw (-.8,-.3)  to (-1.0,-.3);
\draw (-1.0,-.3)  to (-1.0,4.7);
\node at (-1.0,2.3) [fill=white,inner sep=1pt, rotate=-90] {$N$ sub-channels};

\draw[->] (1.1,-1.0) to (1.1,5);
\draw[dotted] (.6,-.3) to (1.3,-.3);
\draw[dotted] (1.7,-.3) to (3,-.3);
\draw[dotted] (.6,.3) to (1.3,.3);
\draw[dotted] (1.7,.3) to (3,.3);
\draw[dotted] (.6,.9) to (3,.9);

\draw[dotted] (.6,2.9) to (1.3,2.9);
\draw[dotted] (.6,3.5) to (1.3,3.5);
\draw[dotted] (.6,4.1) to (3,4.1);
\draw[dotted] (.6,4.7) to (3,4.7);
\node at (1.2,-.3) [right] {$1$};
\node at (1.2,.3) [right] {$\gamma$};
\node at (1.2,.9) [right, fill=white,inner sep=1pt] {$\gamma^2$};
\node at (1.2,2.9) [right] {$\gamma^{N-3}$};
\node at (1.2,3.5) [right] {$\gamma^{N-2}$};
\node at (1.2,4.1) [right, fill=white,inner sep=1pt] {$\gamma^{N-1}$};
\node at (1.2,4.7) [right, fill=white,inner sep=1pt] {$\gamma^{N}=\Gamma$};
\node at (1.1,5.3) [] {\footnotesize{Power-levels}};

\node at (3.2,0) [rectangle, draw, thin, fill=white, minimum width=1.0cm, minimum height=.6cm, rotate=0] {$x_{1}$};
\node at (3.2,.6) [rectangle, draw, thin, fill=white, minimum width=1.0cm, minimum height=.6cm, rotate=0] {$x_{2}$};
\node at (3.2,1.2) [rectangle, draw, thin, fill=white, minimum width=1.0cm, minimum height=.6cm, rotate=0] {$x_{3}$};
\node at (3.2,1.8) [rectangle, draw, thin, fill=white, minimum width=1.0cm, minimum height=.6cm, rotate=0] {$x_{4}$};
\node at (3.2,3.1) [rectangle, draw, thin, fill=white, minimum width=1.0cm, minimum height=2cm, rotate=0] {$\vdots$};
\node at (3.2,4.4) [rectangle, draw, thin, fill=white, minimum width=1.0cm, minimum height=.6cm, rotate=0] {$x_{N}$};
\node at (3.2,-1.2) {Rx};
\node at (3.2,-.6) [rectangle, draw, thin, fill=gray!20, minimum width=1.0cm, minimum height=.6cm, rotate=0] {$z$};
\draw (4.0,4.7) to (4.2,4.7);
\draw (4.0,-.3) to (4.2,-.3);
\draw (4.2,-.3) to (4.2,4.7);
\node at (4.2,2.3) [fill=white,inner sep=1pt, rotate=-90] {$N$ sub-channels};
\end{tikzpicture}
\label{Fig:ChannelDecompP2P}
}\hspace{.7cm}
\subfigure[A $2\times 1$ channel as a set of successive sub-channels. The receiver observes the sum of the signals plus noise.]{
\begin{tikzpicture}[semithick,scale=.93, every node/.style={scale=.93}]]

 \node at (2.7,5.3) [] {\footnotesize{Power levels}};
 \draw[->] (2.7,-1.0) to (2.7,5);
 \node at (2.7,-.3) [right] {$1$};
 \node at (2.7,.3) [right] {$\gamma$};
 \node at (2.7,.9) [right] {$\gamma^2$};
 \node at (2.7,2.9) [right] {$\gamma^{N_1-3}$};
 \node at (2.7,3.5) [right] {$\gamma^{N_1-2}=\Gamma_2$};
 \node at (2.7,4.1) [right] {$\gamma^{N_1-1}$};
 \node at (2.7,4.7) [right] {$\gamma^{N_1}=\Gamma_1$};
 \draw[dotted] (.1,-.3) to (.5,-.3);
 \draw[dotted] (.1,.3) to (.5,.3);
 \draw[dotted] (.1,.9) to (.5,.9);
 \draw[dotted] (.1,2.9) to (.5,2.9);
 \draw[dotted] (.1,3.5) to (.5,3.5);
 \draw[dotted] (.1,4.1) to (2.7,4.1);
 \draw[dotted] (.1,4.7) to (2.7,4.7);
 
 \draw[dotted] (1.9,-.3) to (2.7,-.3);
 \draw[dotted] (1.9,.3) to (2.7,.3);
 \draw[dotted] (1.9,.9) to (2.7,.9);
 \draw[dotted] (1.9,2.9) to (2.7,2.9);
 \draw[dotted] (1.9,3.5) to (2.7,3.5);
 
\node (x1) at (-4.8,0) [rectangle, draw, thin, fill=white, minimum width=1.1cm, minimum height=.6cm, rotate=0] {$x_{1,1}$};
\node (x2) at (-4.8,.6) [rectangle, draw, thin, fill=white, minimum width=1.1cm, minimum height=.6cm, rotate=0] {$x_{1,2}$};
\node (dots) at (-4.8,1.9) [rectangle, draw, thin, fill=white, minimum width=1.1cm, minimum height=2cm, rotate=0] {$\vdots$};
\node (xNm2) at (-4.8,3.2) [rectangle, draw, thin, fill=white, minimum width=1.1cm, minimum height=.6cm, rotate=0] {};
\node at (-4.8,3.2) {\scriptsize{$x_{1,N_1-2}$}};
\node (xNm1) at (-4.8,3.8) [rectangle, draw, thin, fill=white, minimum width=1.1cm, minimum height=.6cm, rotate=0] {};
\node at (-4.8,3.8) {\scriptsize{$x_{1,N_1-1}$}};
\node (xN) at (-4.8,4.4) [rectangle, draw, thin, fill=white, minimum width=1.1cm, minimum height=.6cm, rotate=0] {$x_{1,N_1}$};
\node at (-4.8,-.6) [rectangle, draw, thin, fill=white, minimum width=1.1cm, minimum height=.6cm, rotate=0] {$0$};

\draw (-5.6,4.7) to (-5.8,4.7);
\draw (-5.6,-.3) to (-5.8,-.3);
\draw (-5.8,-.3) to (-5.8,4.7);
\node at (-5.8,2.3) [fill=white,inner sep=1pt, rotate=-90] {$N_1$ sub-channels};
\node at (-4.8,-1.2) {Tx$_1$};

\node (x1) at (-3.3,0) [rectangle, draw, thin, fill=white, minimum width=1.1cm, minimum height=.6cm, rotate=0] {$x_{2,1}$};
\node (x2) at (-3.3,.6) [rectangle, draw, thin, fill=white, minimum width=1.1cm, minimum height=.6cm, rotate=0] {$x_{2,2}$};
\node (dots) at (-3.3,1.9) [rectangle, draw, thin, fill=white, minimum width=1.1cm, minimum height=2cm, rotate=0] {$\vdots$};
\node (xNm2) at (-3.3,3.2) [rectangle, draw, thin, fill=white, minimum width=1.1cm, minimum height=.6cm, rotate=0] {$x_{2,N_2}$};
\node at (-3.3,-.6) [rectangle, draw, thin, fill=white, minimum width=1.1cm, minimum height=.6cm, rotate=0] {$0$};

\draw (-2.5,3.5) to (-2.3,3.5);
\draw (-2.5,-.3) to (-2.3,-.3);
\draw (-2.3,-.3) to (-2.3,3.5);
\node at (-2.3,1.6) [fill=white,inner sep=1pt, rotate=-90] {$N_2$ sub-channels};
\node at (-3.3,-1.2) {Tx$_2$};

\node (x1) at (-.5,0) [rectangle, draw, thin, fill=white, minimum width=1.1cm, minimum height=.6cm, rotate=0] {$x_{1,1}$};
\node (x2) at (-.5,.6) [rectangle, draw, thin, fill=white, minimum width=1.1cm, minimum height=.6cm, rotate=0] {$x_{1,2}$};
\node (dots) at (-.5,1.9) [rectangle, draw, thin, fill=white, minimum width=1.1cm, minimum height=2cm, rotate=0] {$\vdots$};
\node (xNm2) at (-.5,3.2) [rectangle, draw, thin, fill=white, minimum width=1.10cm, minimum height=.6cm, rotate=0] {};
\node at (-.5,3.2) {\scriptsize{$x_{1,N_1-2}$}};
\node (xNm1) at (-.5,3.8) [rectangle, draw, thin, fill=white, minimum width=1.1cm, minimum height=.6cm, rotate=0] {};
\node at (-.5,3.8) {\scriptsize{$x_{1,N_1-1}$}};
\node (xN) at (-.5,4.4) [rectangle, draw, thin, fill=white, minimum width=1.1cm, minimum height=.6cm, rotate=0] {$x_{1,N_1}$};

\draw (-1.2,4.7) to (-1.4,4.7);
\draw (-1.2,-.3)  to (-1.4,-.3);
\draw (-1.4,-.3)  to (-1.4,4.7);
\node at (-1.4,2.3) [fill=white,inner sep=1pt, rotate=-90] {$N_1$ sub-channels};

\node (x1) at (1.2,0) [rectangle, draw, thin, fill=white, minimum width=1.1cm, minimum height=.6cm, rotate=0] {$x_{2,1}$};
\node (x2) at (1.2,.6) [rectangle, draw, thin, fill=white, minimum width=1.1cm, minimum height=.6cm, rotate=0] {$x_{2,2}$};
\node (dots) at (1.2,1.9) [rectangle, draw, thin, fill=white, minimum width=1.1cm, minimum height=2cm, rotate=0] {$\vdots$};
\node (xNm2) at (1.2,3.2) [rectangle, draw, thin, fill=white, minimum width=1.1cm, minimum height=.6cm, rotate=0] {$x_{2,N_2}$};
\node at (.35,2) [] {+};
\draw (2.0,3.5) to (2.2,3.5);
\draw (2.0,-.3) to (2.2,-.3);
\draw (2.2,-.3) to (2.2,3.5);
\node at (2.2,1.6) [fill=white,inner sep=1pt, rotate=-90] {$N_2$ sub-channels};

\node at (0.35,-.6) [rectangle, draw, thin, fill=gray!20, minimum width=2.8cm, minimum height=.6cm, rotate=0] {$z$};
\node at (0.35,-1.2) {Rx};
\end{tikzpicture}
\label{Fig:ChannelDecompMAC}}
\caption{SCD for a P2P channel and a Many-to-One channel.}
\end{figure*}

We interpret each of the $N$ power-levels as a sub-channel (Fig. \ref{Fig:ChannelDecompP2P}), with an $\snr$ of $\gamma$. Since these sub-channels are accessed successively, we call them {\it `successive sub-channels'}. Next, we extend this SCD to two elemental multi-user channels: many-to-one and one-to-many channels.

\subsection{Many-to-one Channels}
\label{SubSec:ManyToOne}
Consider a Gaussian channel with inputs $X_1$ and $X_2$ with power constraints $\Gamma_1$ and $\Gamma_2\leq \Gamma_1$, respectively, one output $Y=X_1+X_2+Z$, and noise $Z\sim\mathcal{N}(0,1)$, i.i.d. over time. Setting $\gamma^{N_1}=\Gamma_1$ with $N_1\in\mathbb{N}\setminus\{0\}$, we describe the channel as a set of $N_1$ successive sub-channels similar to the P2P channel. 

Now, we decompose the second transmit signal into $N_2$ signals $x_2=\sum_{\ell=1}^{N_2} x_{2,\ell}$, where $x_{2,\ell}$ has power $p_\ell=\gamma^\ell-\gamma^{\ell-1}$. The power levels of the $N_2$ signals align with the first $N_2$ power-levels of $x_1$ (Fig. \ref{Fig:ChannelDecompMAC}). Here, $N_2$ is chosen as $N_2=\left\lfloor\frac{\log(\Gamma_2)}{\log(\gamma)}\right\rfloor$. 

The achievable rate over each sub-channel depends on two parameters: the per sub-channel $\snr$ $\gamma$, and the way the sub-channels are used. Namely, a sub-channel can be exclusively used by one user, or shared among multiple users.

\subsubsection{One user per sub-channel}
This can be the case in a multiple-access channel (MAC), where sub-channels are allocated exclusively to users. If we denote the number of sub-channels allocated to user $i$ by $r_i$, then we must have $r_2\leq N_2$ and $r_1+r_2\leq N_1$. Note that sub-channels $N_2+1,\cdots,N_1$ can only be used by user 1. The achievable rate region can be calculated by multiplying $r_i$ by the achievable rate per sub-channel $\hat{C}(\gamma)$. As a result, the region
\begin{align}
R_2&\leq N_2\hat{C}(\gamma)\approx\hat{C}(\Gamma_2)\\
R_1+R_2&\leq N_1\hat{C}(\gamma)=\hat{C}(\Gamma_1),
\end{align}
is achievable.\footnote{Here, we have approximated $N_2=\left\lfloor \frac{\log(\Gamma_2)}{\log(\gamma)}\right\rfloor$ by $\frac{\log(\Gamma_2)}{\log(\gamma)}$, which can be made precise by choosing $N_1$ appropriately.} This region is within a constant gap of the capacity region of the MAC and provides a good approximation.

\subsubsection{Two users per sub-channel}
If two users are allowed to share a sub-channel\footnote{This   is of particular interest for problems of computation over multi-user channels \cite{NazerGastpar} and for interference networks \cite{SridharanJafarianVishwanathJafar}.}, then the rate per sub-channel becomes $\hat{C}(\gamma/2)$. To show this, denote by $\mathcal{I}_i$ the set of sub-channels allocated to user $i$. Note that $\mathcal{I}_{12}=\mathcal{I}_1\cap\mathcal{I}_2\neq\emptyset$ due to sub-channel sharing. Then, the interference plus noise power experienced by the receiver  while decoding the signal on sub-channel $\ell>N_2$ is
\begin{align}
1+\sum_{\substack{i=1\\i\in\mathcal{I}_1}}^{\ell-1}p_i
+\sum_{\substack{i=1\\i\in\mathcal{I}_2}}^{\ell-1}p_i
&< 1+\sum_{i=1}^{\ell-1}2p_i
< 2\gamma^{\ell-1}.
\end{align}
Thus, while decoding the signal on the $\ell$-th sub-channel, the $\snr$ is at least $\gamma/2$ leading to an achievable rate per sub-channel of $\hat{C}(\gamma/2)$. This rate is also achievable for computation using lattice codes \cite{NazerGastpar} over sub-channel $\ell\leq N_2$. Namely, if the receiver wants to recover $x_{1,\ell}+x_{2,\ell}$, then the rates of $x_{1,\ell}$ and $x_{2,\ell}$ denoted $R_{1,\ell}$ and $R_{2,\ell}$ should satisfy
\begin{align}
R_{1,\ell},R_{2,\ell}\leq\hat{C}\left(\frac{1}{2}+\frac{\gamma^\ell-\gamma^{\ell-1}}{2\gamma^{\ell-1}}\right)=\hat{C}\left(\frac{\gamma}{2}\right).
\end{align}


Generalizing this to $K$ users leads to the following lemma.

\begin{lemma}[SCD of many-to-one channels]
\label{Lem:ManyToOne}
A $K\times 1$ Gaussian channel with $\snr$'s of \ $\Gamma_1\geq \Gamma_2\geq\cdots\geq\Gamma_K$ can be decomposed into a set of $N_1\in\mathbb{N}\setminus\{0\}$ successive sub-channels where user $i$ has access to sub-channels $1,\cdots,N_i$, with $N_i=\left\lfloor\frac{\log(\Gamma_i)}{\log(\gamma)}\right\rfloor$ for $i=2,\cdots,K$ and $\gamma^{N_1}=\Gamma_1$. The decoding/computation rate of each sub-channel is $\hat{C}(\gamma/\kappa)$ where $\kappa=\max_{\ell}\kappa_\ell$ and $\kappa_\ell$ is the number of users using sub-channel $\ell$.
\end{lemma}

\begin{remark}
This decomposition is only meaningful for $\Gamma_1>\kappa^{N_1}$ since $\hat{C}\left(\gamma/\kappa\right)<0$ otherwise.
\end{remark}

This lemma simplifies the problem of rate achievability over a Gaussian many-to-one channel to a sub-channel allocation problem. 
Note that the per sub-channel rate reduction for $\kappa>1$ in comparison with $\kappa=1$ is $\hat{C}(\kappa)$, for a total rate reduction of at most $N_1\hat{C}(\kappa)$. This reduction is however independent of $\Gamma_1,\cdots,\Gamma_K$, and thus yields a constant gap as a function of $\Gamma_i$.

\subsection{One-to-many Channels}
\label{SubSec:OneToMany}
Consider a Gaussian channel with input $X$ with a power constraint $\Gamma$, and two outputs $Y_i=X+Z_i$, $i=1,2$, where $Z_i$ is $\mathcal{N}(0,\sigma_i^2)$, i.i.d. over time. Further, let $\sigma_2^2\leq \sigma_1^2=1$ and denote the $\snr$ of $Y_i$ as $\Gamma_i=\Gamma/\sigma_i^2$. We start by decomposing the channel to receiver 1 into a set of $N_1$ successive sub-channels, with $\gamma^{N_1}=\Gamma_1$ similar to above. 

\begin{figure}
\centering
\begin{tikzpicture}[semithick]
\node at (5.6,5.3) [] {\footnotesize{Power levels}};
\draw[->] (5.6,-1.0) to (5.6,5);
\draw[dotted] (5.7,-.3) to (3,-.3);	
\draw[dotted] (5.7,.3) to (3,.3);
\draw[dotted] (6.35,.6) to (4.4,.6);
\draw[dotted] (5.7,.9) to (3,.9);
\draw[dotted] (5.7,1.5) to (3,1.5);
\draw[dotted] (5.7,2.1) to (3,2.1);
\draw[dotted] (5.7,4.1) to (3,4.1);
\draw[dotted] (5.7,4.7) to (3,4.7);	
\node at (5.7,-.3) [right] {$1$};
\node at (5.7,.3) [right] {$\gamma$};
\node at (6.4,.6) [right, fill=white,inner sep=1pt] {$\sigma_2^2$};
\node at (5.7,.9) [right, fill=white,inner sep=1pt] {$\gamma^2$};
\node at (5.7,1.5) [right, fill=white,inner sep=1pt] {$\gamma^3=\gamma^q$};
\node at (5.7,2.1) [right, fill=white,inner sep=1pt] {$\gamma^4$};
\node at (5.7,4.1) [right, fill=white,inner sep=1pt] {$\gamma^{N_1-1}$};
\node at (5.7,4.7) [right, fill=white,inner sep=1pt] {$\gamma^{N_1}=\Gamma_1$};

\node (x1) at (0,0) [rectangle, draw, thin, fill=white, minimum width=1.1cm, minimum height=.6cm, rotate=0] {$x_{1}$};
\node (x2) at (0,.6) [rectangle, draw, thin, fill=white, minimum width=1.1cm, minimum height=.6cm, rotate=0] {$x_{2}$};
\node (dots) at (0,1.9) [rectangle, draw, thin, fill=white, minimum width=1.1cm, minimum height=2cm, rotate=0] {$\vdots$};
\node (xNm2) at (0,3.2) [rectangle, draw, thin, fill=white, minimum width=1.1cm, minimum height=.6cm, rotate=0] {\scriptsize{$x_{N_1-2}$}};
\node (xNm1) at (0,3.8) [rectangle, draw, thin, fill=white, minimum width=1.1cm, minimum height=.6cm, rotate=0] {\scriptsize{$x_{N_1-1}$}};
\node (xN) at (0,4.4) [rectangle, draw, thin, fill=white, minimum width=1.1cm, minimum height=.6cm, rotate=0] {$x_{N_1}$};

\node at (0,-.6) [rectangle, draw, thin, fill=white, minimum width=1.1cm, minimum height=.6cm, rotate=0] {$0$};

\node at (0,-1.2) {Tx};
\draw (-0.7,4.7) to (-.9,4.7);
\draw (-0.7,-.29) to (-.9,-.29);
\draw (-0.9,-.29) to (-.9,4.7);
\node at (-.9,2.1) [fill=white,inner sep=1pt, rotate=-90] {$N_1$ sub-channels};
\draw (0.7,4.7) to (.9,4.7);
\draw (0.7,.9) to (.9,.9);
\draw (.9,.9) to (.9,4.7);
\node at (.9,2.8) [fill=white,inner sep=1pt, rotate=-90] {$N_2$ sub-channels};

\node at (2.6,0) [rectangle, draw, thin, fill=white, minimum width=1.1cm, minimum height=.6cm, rotate=0] {$x_{1}$};
\node at (2.6,.6) [rectangle, draw, thin, fill=white, minimum width=1.1cm, minimum height=.6cm, rotate=0] {$x_{2}$};
\node at (2.6,1.2) [rectangle, draw, thin, fill=white, minimum width=1.1cm, minimum height=.6cm, rotate=0] {$x_{3}$};
\node at (2.6,1.8) [rectangle, draw, thin, fill=white, minimum width=1.1cm, minimum height=.6cm, rotate=0] {$x_{4}$};
\node at (2.6,3.1) [rectangle, draw, thin, fill=white, minimum width=1.1cm, minimum height=2cm, rotate=0] {$\vdots$};
\node at (2.6,4.4) [rectangle, draw, thin, fill=white, minimum width=1.1cm, minimum height=.6cm, rotate=0] {$x_{N_1}$};
\node at (2.6,-.6) [rectangle, draw, thin, fill=gray!20, minimum width=1.1cm, minimum height=.6cm, rotate=0] {$z_1$};
\draw (1.9,4.7) to (1.7,4.7);
\draw (1.9,-.29) to (1.7,-.29);
\draw (1.7,-.29) to (1.7,4.7);
\node at (1.7,2.1) [fill=white,inner sep=1pt, rotate=-90] {$N_1$ sub-channels};
\node at (2.65,-1.2) {Rx$_1$};

\node at (4.2,.6) [rectangle, draw, thin, fill=white, minimum width=1.1cm, minimum height=.6cm, rotate=0] {$ $};
\node (x1) at (4.2,1.2) [rectangle, draw, thin, fill=white, minimum width=1.1cm, minimum height=.6cm, rotate=0] {$x_{3}$};
\node (x2) at (4.2,1.8) [rectangle, draw, thin, fill=white, minimum width=1.1cm, minimum height=.6cm, rotate=0] {$x_{4}$};
\node (dots) at (4.2,3.1) [rectangle, draw, thin, fill=white, minimum width=1.1cm, minimum height=2cm, rotate=0] {$\vdots$};
\node (xNm2) at (4.2,4.4) [rectangle, draw, thin, fill=white, minimum width=1.1cm, minimum height=.6cm, rotate=0] {$x_{N_1}$};
\node at (4.2,-0.15) [rectangle, draw, thin, fill=gray!20, minimum width=1.1cm, minimum height=1.5cm, rotate=0] {$z_2$};
\draw (4.9,4.7) to (5.1,4.7);
\draw (4.9,.9) to (5.1,.9);
\draw (5.1,.9) to (5.1,4.7);
\node at (5.1,2.8) [fill=white,inner sep=1pt, rotate=-90] {$N_2$ sub-channels};
\node at (4.25,-1.2) {Rx$_2$};

\end{tikzpicture}
\caption{A $1\times2$ channel as a set of successive sub-channels. Rx$_1$ observes all sub-channels, while the more noisy Rx$_2$ observes only $N_2$ sub-channels since the lower sub-channels have power-levels smaller than noise power.}
\label{Fig:ChannelDecompBC}
\end{figure}
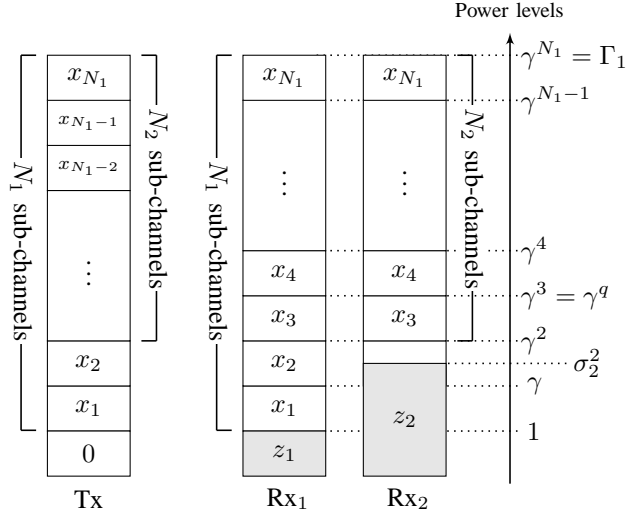

Receiver 2 receives some of the transmit signals sent over the sub-channels at a power lower than noise. When decoding the signal over sub-channel $q$ (with $\gamma^q>\sigma_2^2$), the interference plus noise power is $\gamma^{q-1}-1+\sigma_2^2$ (instead of $\gamma^{q-1}$ at receiver 1). We would like to bound $q$ so that all sub-channels $q,\cdots,N_1$ can supply a rate within a constant of $\hat{C}(\gamma)$ at receiver 2. To this end, let $q$ be the smallest integer such that 
\begin{align}
\label{Eq:ConditionQ}
\gamma^{q-1}>\sigma_2^2-1.
\end{align}
For decoding a signal on sub-channel $q$ at receiver 2, the following rate can be supported
\begin{align}
C\left(\frac{\gamma^q-\gamma^{q-1}}{\gamma^{q-1}-1+\sigma_2^2}\right)
&> C\left(\frac{\gamma^q-\gamma^{q-1}}{2\gamma^{q-1}}\right)
> \hat{C}\left(\frac{\gamma}{2}\right). \label{Eq:C_subchannel_q}
\end{align}
The same holds for sub-channels $q+1,\cdots,N_1$. The smallest integer $q$ that satisfies the condition \eqref{Eq:ConditionQ} is given by
\begin{align}
q=\left\lceil\frac{\log(\sigma_2^2-1)}{\log(\gamma)}+1\right\rceil.
\end{align}
The number of sub-channels that are accessible by receiver 2 (with power-level above noise) is $N_1+1-q$, which is lower bounded by
\begin{align}
N_1+1-q&>N_1+1-\left\lceil\frac{\log(\sigma_2^2)}{\log(\gamma)}+1\right\rceil\\
&=N_1+1-\left\lceil N_1-\frac{\log(\Gamma_2)}{\log(\gamma)}+1\right\rceil\\
&\geq N_1+1-\left\lceil N_1-\left\lfloor\frac{\log(\Gamma_2)}{\log(\gamma)}\right\rfloor+1\right\rceil\\
&=N_2,
\end{align}
where $N_2=\left\lfloor\frac{\log(\Gamma_2)}{\log(\gamma)}\right\rfloor$ (see Fig. \ref{Fig:ChannelDecompBC}).

Note that the factor $\frac{1}{2}$ in \eqref{Eq:C_subchannel_q} arises due to decoding signals at the weaker receiver, where the impact of noise is `doubled' due to interference. By taking this effect into account, we can state the following lemma.

\begin{lemma}[SCD of one-to-many channels]
\label{Lem:OneToMany}
A $1\times K$ Gaussian channel with $\snr$'s of $\Gamma_1\geq \Gamma_2\geq\cdots\geq \Gamma_K$ can be decomposed into a set of $N_1\in\mathbb{N}\setminus\{0\}$ successive sub-channels, where user $i$ has access to sub-channels $N_1-N_i+1,\cdots,N_1$, with $N_i=\left\lfloor\frac{\log(\Gamma_i)}{\log(\gamma)}\right\rfloor$ for $i=2,\cdots,K$ and $\gamma^{N_1}=\Gamma_1$. The achievable-rate of each sub-channel is $\hat{C}(\gamma/(1+\mu))$ where $\mu=0$ if $K=1$ and $\mu=1$ otherwise.
\end{lemma}

\begin{remark}
This decomposition is only meaningful if $\Gamma_1>2^{N_1}$ since $\hat{C}\left(\gamma/2\right)<0$ otherwise, 
\end{remark}

\subsection{Main Communication Strategies over the Sub-Channels}
\label{Sec:CommStrategies}
We are interested in 3 strategies that can be used over these sub-channels, namely: decoding, computation, and neutralization. The achievable rate per sub-channel can be generally written as 
\begin{align}
\label{RperSubChannel}
\hat{C}(\gamma/(\kappa+\mu))
\end{align}
where $\kappa$ is as defined in Lemma \ref{Lem:ManyToOne} and $\mu=0$ if the network has one receiver and $\mu=1$ otherwise (Lemma \ref{Lem:OneToMany}).

Decoding refers to the case where a signal is desired by only one receiver. It is described similar to the P2P description in the beginning of this Appendix. We describe computation next.

\subsubsection{Computation}
Assume that users $1,\cdots,\kappa'\leq \kappa$, transmit over sub-channel $\ell$ of a many-to-one channel. A function of their transmit codewords has to be computed at the receiver. User $i\in\{1,\cdots,\kappa'\}$ uses an $n$-dimensional nested-lattice codebook with fine lattice $\Lambda_f$, coarse lattice $\Lambda_c$, rate $R_i$, and power 1 (see \cite{NazerGastpar} for more details on lattice codes). The transmit signal of user $i=1,\cdots,\kappa'$ over sub-channel $\ell$ is
\begin{align}
x_{i,\ell}^{(n)}=\sqrt{p_\ell}[(\lambda_i+d_i)\bmod\Lambda_c],
\end{align}
where $\lambda_1$ is the nested-lattice codeword, $d_1$ is a random dither, and $p_\ell$ is the power of sub-channel $\ell$.  The receiver observes
\begin{align}
y_\ell^{(n)}=\sum_{i=1}^{\kappa'}x_{i,\ell}^{(n)}+z_\ell^{(n)},
\end{align}
over this sub-channel, where $z_\ell^{(n)}$ is a length-$n$ sequence that contains the interference plus noise signals (with power $\kappa\gamma^{\ell-1}$ at most). The receiver computes the modulo-sum $S(\lambda_1,\cdots,\lambda_{\kappa'})=\left(\sum_{i=1}^{\kappa'}\lambda_i\right)\bmod\Lambda_c$, which is possible as long as \cite{NazerGastpar}
\begin{align}
R_i\leq R_\ell=\hat{C}\left(\frac{1}{\kappa'}+\frac{\gamma}{\kappa}\right),\quad i=1,\cdots,\kappa'.
\end{align}
After computing this modulo-sum, the receiver reconstructs $\sum_{i=1}^{\kappa'}x_{i,\ell}^{(n)}$ \cite{Nazer_IZS2012} and  subtracts it from the received signal in order to proceed with decoding the remaining sub-channels.

If the sub-channel has multiple receivers, then $\mu=1$ and the achievable rate becomes
$$R_\ell=\hat{C}\left(\frac{1}{\kappa'}+\frac{\gamma}{\kappa+1}\right)>\hat{C}\left(\frac{\gamma}{\kappa+1}\right),$$
which is consistent with \eqref{RperSubChannel}.

\subsubsection{Neutralization}
\label{SubSubSection:Neutralization}
Consider a setup with two transmitters and one receiver, and let $\lambda_1$ and $\lambda_2$ be nested-lattice codewords with rates $R$ and unit power. Transmitter $1$ wants to send a codeword $\lambda_1$ to the receiver, and has knowledge of $\lambda_0=\left(\lambda_1-\lambda_2\right)\bmod\Lambda_c$ which also has rate $R$. Transmitter $2$ has knowledge of $\lambda_2$. 
 
The two transmitters cooperate to send $\lambda_1$ to the receiver over sub-channel $\ell$ as follows. The transmitters send
\begin{align}
x_{1,\ell}^{(n)}&=\sqrt{p_\ell}[(\lambda_0+d_0)\bmod\Lambda_c],\\
x_{2,\ell}^{(n)}&=\sqrt{p_\ell}[(\lambda_2+d_2)\bmod\Lambda_c],
\end{align}
respectively, where $d_0$ and $d_2$ are random dithers. The receiver obtains
\begin{align}
y_\ell^{(n)}=x_{1,\ell}^{(n)}+x_{2,\ell}^{(n)}+z_\ell^{(n)}.
\end{align}
Now the receiver computes $(\lambda_0+\lambda_2)\bmod\Lambda_c$. This is possible as long as the rates satisfy
\begin{align}
R\leq \hat{C}\left(\frac{1}{2}+\frac{\gamma}{\kappa+\mu}\right),
\end{align}
where $\kappa$ is the maximum number of allowed users per sub-channel, and $\mu$ is zero if the channel has one receiver and one otherwise. Then the receiver recovers $(\lambda_0+\lambda_2)\bmod\Lambda_c=(\lambda_1)\bmod\Lambda_c=\lambda_1$, and the interference from $\lambda_2$ is neutralized  \cite{MohajerDiggaviFragouliTse_Neutralization,ChaabanSezginTuninetti_Asilomar}. Therefore, the rate
\begin{align}
R_\ell=\hat{C}(\gamma/(\kappa+\mu))
\end{align}
is also achievable for neutralization, which is also consistent with \eqref{RperSubChannel}.

\section{Sub-channel Grouping}
\label{SubSec:SubChannelGrouping}
The achievable rate region obtained using the SCD has a `gap' term which depends on the number of decoding steps involved. Namely, each decoding step incurs a gap of $\frac{1}{2}\log(\kappa+\mu)$ bit. This gap can be reduces by reducing the number of decoding steps involved in the scheme.

To achieve this goal, we note that several sub-channels might be used similarly in a given scheme. Consider the Y-channel as an example. In the Y-channel, sub-channels used for bi-directional communication between the user pair $(1,2)$ are used similarly. Similar comment applies to user pairs $(1,3)$ and $(2,3)$. Thus, we have 3 usage cases for bi-directional communication. For cyclic communication, we have four such cases (see Sec. \ref{Sec:YCAchievability}). For uni-directional communication, we have six cases, one per each direction $i\to j$. This leads to a total of 13 different usage cases.

To reduce the number of decoding steps, we can allocate consecutive sub-channels to each usage case, then combine these sub-channels. Consider two sub-channels $\ell$ and $\ell-1$, and the signals sent over those sub-channels with powers $\gamma^\ell-\gamma^{\ell-1}$ and $\gamma^{\ell-1}-\gamma^{\ell-2}$ and rates $R_\ell=R_{\ell-1}$. Computing the modulo-sum of the signals at the receiver can be achieved reliably if $R_\ell=R_{\ell-1}\leq\hat{C}(\gamma/2)$. Now let each of the two transmitters using those sub-channels combine its two signals into one, with power $(\gamma^\ell-\gamma^{\ell-1})+(\gamma^{\ell-1}-\gamma^{\ell-2})
=\gamma^\ell-\gamma^{\ell-2}$ and rate $R$, and send this signal instead. This modulo-sum of the signals can be decoded (computation) reliably if
\begin{align}
R&\leq \hat{C}\left(\frac{1}{2}+\frac{\gamma^\ell-\gamma^{\ell-2}}{2\gamma^{\ell-2}}\right)
= 2\hat{C}\left(\gamma\right)-\frac{1}{2}.
\end{align}
The resulting rate constraint is equal to the sum of the rate constraints on $R_\ell$ and $R_{\ell-1}$ plus $\frac{1}{2}$, thus reducing the gap by $\frac{1}{2}$.

This idea can be generalized to all signals and sub-channels of the Y-channel. The number of sub-channel groups will be 13 in the Y-channel. Since the bounds \eqref{ARCYC1}-\eqref{ARCYC8} bound combinations of 2 and 3 rates, and since each rate is split into 3 parts (bi-directional, cyclic, and uni-directional), each such rate constraint corresponds to decoding on either 6 or 9 sub-channel groups, respectively. This leads to a gap of at most $\frac{6}{2}$ and $\frac{9}{2}$ bits, respectively. Thus, the gaps $\widetilde{N}_3/2$ and $\widetilde{N}_2/2$ in \eqref{ARCYC1}-\eqref{ARCYC8} can be replaced with 3 and 9/2, respectively. This achievable rate region is less than 2 bits per dimension away from the outer bound in \cite[Theorem 5]{ChaabanSezgin_YC_Reg}. 

Note that a more involved study dedicated for the Y-channel  leads to a smaller gap at the expense of a more difficult analysis \cite{ChaabanSezgin_YC_Reg}.

\end{appendices}

\bibliography{myBib}

\end{document}